\documentclass[aps,amsmath,amssymb,showpacs,showkeys]{revtex4-2}
\usepackage[dvips]{graphicx,color}
\usepackage{times}
\usepackage{xcolor, soul}
\sethlcolor{yellow}
\usepackage[
  colorlinks=true,
  urlcolor=blue,
  linkcolor=red,
  citecolor=blue
]{hyperref}
\usepackage{bm}
\usepackage{amsfonts}
\usepackage{latexsym}
\usepackage[latin1]{inputenc}
\usepackage{graphicx}
\usepackage{amsmath}
\usepackage{palatino}
\usepackage{mathpazo}
\usepackage{textcomp}
\usepackage{float}
\usepackage{booktabs}
\usepackage{dcolumn}
\usepackage{ragged2e}
\usepackage{hyperref}
\hypersetup{colorlinks,citecolor=blue}
\hypersetup{colorlinks=true, linkcolor=red, filecolor=magenta, urlcolor=blue}
\usepackage{amsmath}
\usepackage{xcolor}
\usepackage{orcidlink}
\usepackage{epsfig}
\usepackage[caption=false]{subfig}
\usepackage{commath}
\usepackage{multirow}
\usepackage{rotating}
\usepackage{tabularx}
\begin{document}

\title{Anisotropic cosmology in bumblebee gravity theory}

\author{Pranjal Sarmah \orcidlink{0000-0002-0008-7228}}
\email[E-mail:]{p.sarmah97@gmail.com}

\author{Umananda Dev Goswami \orcidlink{0000-0003-0012-7549}}
\email[E-mail:]{umananda@dibru.ac.in}

\affiliation{Department of Physics, Dibrugarh University, Dibrugarh 786004, 
Assam, India}

%\date{}
\begin{abstract}
The bumblebee vector model of spontaneous Lorentz symmetry breaking (LSB) in 
Bianchi type I (BI) Universe to observe its effect on cosmological evolution 
is an interesting aspect of study in anisotropic cosmology. In this study, 
we have considered a bumblebee field under vacuum expectation value condition 
(VEV) with BI metric and studied the cosmological parameters along with 
observational data. Further, we have studied the effect of anisotropy and 
the bumblebee field in cosmic evolution. We have also studied the effect of 
both anisotropy and bumblebee field while considering the Universe as a 
dynamical system. We have found that there are some prominent roles of both 
anisotropy and the bumblebee field in cosmic evolution. We have also observed 
an elongated matter-dominated phase as compared to standard cosmology. 
Moreover, while studying the dynamical system analysis, we have also observed 
the shift of critical points from standard $\Lambda$CDM results showing the 
anisotropy and the bumblebee field effect.
 
\end{abstract}

%\pacs{04.30.Tv, 04.50.Kd}

\maketitle
\section{Introduction}\label{1}
Standard cosmology, also known as $\Lambda$CDM cosmology mainly relies on the 
two cosmological principles, viz., the isotropy and the homogeneity of the 
Universe on a large scale. Along with the Friedmann-Lema\^itre-Robertson-Walker
(FLRW) metric supported by an energy-momentum tensor of conventional perfect 
fluid form, this theory provides answers to many questions regarding the 
understanding of the Universe \cite{Pebbles_1994}. However, the observed 
accelerated expansion of the Universe \cite{Riess, Perlmutter_1999, Ma_2011} 
along with the absence of observational evidence on the dark matter (DM)
\cite{Trimble_87} and dark energy (DE) \cite{Frieman_2008} etc.~have motivated
researchers to search for alternate theories or modifications in general 
relativity (GR) through expanding the conventional formalism 
\cite{Capelo_2015}. In this context, a gravitational model known as the 
bumblebee gravity model was proposed \cite{Kostelecky_1989}. This model is 
beyond the ambit of GR, which relaxes the cosmological principles. In this 
theory, a vector field known as the bumblebee field has been introduced and
as a result, it modifies the Einstein field equations of GR.

The bumblebee model was first introduced in 1989 by Kostelecky and Samuel 
\cite{Kostelecky_1989}. This is a simple yet reliable standard model extension 
(SME) that basically works in the principle of Lorentz symmetry breaking (LSB) 
\cite{Kostelecky_1989, Kostelecky_1989a} through using a vector field. The 
model introduces this field and the potential in the conventional 
Einstein-Hilbert (EH) action which results in the modification of the general 
Einstein field equations and it helps to understand the various cosmological
aspects without considering the exotic matter and energy contents like DM and 
DE in the Universe \citep{Capelo_2015, Bertolami_2005}. However, several other 
SMEs are also available, in which the EH action was modified through the 
inclusion of some nontrivial curvature terms 
\cite{Felice_2010, Bertolami_2014}, suitable scalar terms \cite{Zlatev_1999}, 
vector fields \cite{Tartagila_2007, Armendariz_2009,Kostelecky_2004} and 
so on. The impact of SMEs in the research of the gravitational sector can 
be found in the Refs.~\cite{Bluhm_2005,Bluhm_2008,Kostelecky_2009, 
Kostelecky_2005a,Kostelecky_2005, Gogoi_2022, Karmakar_2023}.

Although, as mentioned already, the isotropic and homogeneous cosmology 
successfully explains most of its aspects with the help of the 
standard $\Lambda$CDM model which includes the Hubble tension 
\cite{Planck_2018, Nedelco_2021}, the $\sigma_8$ tension \cite{Planck_2018},
the coincidence problem \cite{Velten_2014} etc., several dependable
observational data sources, including WMAP \cite{wmap,wmap1,wmap2},
SDSS (BAO) \cite{SDSS_2005,Bessett_2009,Tully_2023}, and Planck 
\cite{ Planck_2015,Planck_2018} have shown some deviations from the principles 
of standard cosmology. These suggest that the Universe may have some 
anisotropies. Additionally, some studies have pointed to a large-scale planar 
symmetric geometry of the Universe, and for such a geometry the eccentricity 
of the order $10^{-2}$ can match the quadrupole amplitude to the observational 
evidence without changing the higher-order multipole of the temperature 
anisotropy in the CMB angular power spectrum \cite{Tedesco_2006}. Again, the 
existence of asymmetry axes in the vast scale geometry of the Universe is also
evident from the polarization analysis of electromagnetic radiation which
is traveling across great cosmological distances \cite{akarsu_2010}. Thus,
the isotropy and homogeneity assumptions are not sufficient to explain all 
the cosmological phenomena completely. To explain these anisotropies, we 
require a metric with a homogeneous background that possesses the anisotropic 
character. Such a class of metrics had already been provided by Luigi Bianchi 
and out of his eleven types of metrics, Bianchi type I (BI) is the simplest yet 
suitable to explain the anisotropy of the Universe \cite{Sarmah_2022}. 
Researchers have already used this metric along with its special case like the 
locally rotationally symmetric BI (LRS-BI) metric to explain different aspects 
of anisotropic cosmology in both GR and other modified theories and some of 
them are found in the Refs.~\cite{Cea_2022, Perivolaropoulos_2014, Berera_2004, Campanelli_2006, Campanelli_2007, Paul_2008, Barrow_1997, Hossienkhani_2018,akarsu_2019,Sarmah_2022,Sarmah_2023, Sarmah_2024,espo}. 

The SME theory of the bumblebee vector field has been extensively studied by 
researchers in the black hole physics \cite{Gogoi_2022, Karmakar_2023, 
Mai_2023, Gu_2022, Saka_2023}. However, there are very limited studies that 
have been carried out on cosmological scenarios. Cosmological implications 
of bumblebee theory on isotropic cosmology can be found in the 
Refs.~\cite{Capelo_2015,Tartagila_2007}. In anisotropic cosmology too the 
study of bumblebee gravity is in the very preliminary stage. In 
Ref.~\cite{Maluf_2021}, the bumblebee field is considered as a source of 
cosmological anisotropies. Another work of bumblebee gravity on the {Kasner} 
metric has been found in Ref.~\cite{Neves_2023}. However, constraining the 
model parameters with the help of various observational data and extensive 
studies of cosmological parameters in bumblebee gravity with Bianchi type I 
metric is an interesting topic of study to understand the Universe as well as 
the role of anisotropy and bumblebee field in cosmic expansion and evolution. 

In this work, we have considered the BI metric along with the bumblebee vector 
model to explain the anisotropic characteristics of the Universe by studying 
the cosmological parameters. Here we have used available observational data 
like Hubble data, Pantheon data, BAO data, etc.~to understand the situation in 
a more realistic and physical way. Further, we have covered the study of the 
effect of anisotropy and bumblebee field on cosmic evolution through the 
study of the evolution of the density parameters against cosmological 
redshift. Finally, we have also studied the dynamical system analysis for the 
considered case to understand the property of the critical points and 
sequences of the evolution of the various phases of the Universe. In all our 
analyses, we have compared our results with standard $\Lambda$CDM results to 
understand the effect of anisotropy and bumblebee field in cosmic expansion.

%%%%%%%%%%%%%%%%%%%%%%%%%%%%%%%%%%%%%%%%%%%%%%%%%%%%%%%%%%%%%%%%%%%%%%%%%%%%
The current article is organized as follows. Starting from this introduction 
part in Section \ref{1}, we have discussed the general form of field equations 
in the bumblebee field in Section \ref{2}. In Section \ref{3}, we have 
developed the field equations and continuity equation for the Bianchi type I 
metric with the consideration of the time-dependent bumblebee field. In 
Section \ref{4}, we have further simplified the situation by considering the 
vacuum expectation value (VEV) condition and derived the field equations and 
cosmological parameters. In Section \ref{5}, we have constrained the model 
parameters by using the techniques of Bayesian inference by using various 
observational data compilations and compared our model results with standard 
cosmology by using the constrained values of the parameters. In Section 
\ref{6}, we have studied the effect of anisotropy and bumblebee field on
cosmological evolutions. In Section \ref{7} we have made the dynamical system
analysis for both standard $\Lambda$CDM and anisotropic bumblebee model. 
Finally, the article has been summarised with conclusions in Section \ref{8}. 
%%%%%%%%%%%%%%%%%%%%%%%%%%%%%%%%%%%%%%%%%%%%%%%%%%%%%%%%%%%%%%%%%%%%%%%%%%%%
\section{The bumblebee gravity and field equations}\label{2}

The bumblebee vector field model and its associated gravity theory are based 
on the principle of spontaneous Lorentz symmetry breaking (LSB) within the 
gravitational context. These models contain a potential term $V$ that, for the 
field configurations, results in nontrivial VEVs. This can have an impact on 
other fields' dynamics that are coupled to the bumblebee field, while also 
maintaining geometric structures and conservation laws that are compatible 
with the standard pseudo-Riemannian manifold used in GR 
\cite{Kostelecky_2004,Bluhm_2005}. The simplest model with a single bumblebee 
vector field $B_\mu$ coupled to gravity in a non-torsional spacetime can be 
described by the action \cite{Capelo_2015,Maluf_2021},
\begin{equation}\label{action}
{S_{B} = \int\!  \sqrt{-g}\left[\frac{1}{2\kappa}(R+\xi B^{\mu}B^{\nu}R_{\mu\nu})- \frac{1}{4}B^{\mu\nu}B_{\mu\nu}-V(B^{\mu}B_{\mu}\pm b^{2})+\mathcal{L_M}\right]d^{4}x,}
\end{equation}
where $\kappa = 8\pi G$, $\xi$ is the coupling constant with the dimension of
$[M^{-2}]$ accounting for the interaction between the bumblebee field and Ricci 
tensor of spacetime, $B^{\mu}$ is the bumblebee vector field, 
$B_{\mu\nu} = \partial_{\mu}B_{\mu}-\partial_{\nu}B_{\mu}$ is the field 
strength tensor, and $\mathcal{L_M}$ is the matter Lagrangian density. 
Further, $b^{2} \equiv b^{\mu}b_{\mu} = \langle B^{\mu}B_{\mu}\rangle_{0} 
\neq 0$ is the expectation value for the contracted bumblebee vector field and 
$V$ is a field potential satisfying the condition: 
$B_{\mu}B^{\nu}\pm b^{2} = 0$. The field equations obtained through varying the
action \eqref{action} with respect to the  metric $g_{\mu\nu}$ can be written 
as
\begin{align}\label{FE}
G_{\mu \nu} = \kappa \Big[ 2V'B_{\mu} B_{\nu} & + B_{\mu\alpha} B^{\alpha}_{~\nu} - \left( V + \frac{1}{4} B_{\alpha \beta} B^{\alpha \beta} \right) g_{\mu\nu}\Big]
+ \xi \left[\frac{1}{2}B^{\alpha} B^{\beta} R_{\alpha\beta}  g_{\mu\nu} - B_\mu B^\alpha R_{\alpha \nu} - B_{\nu}B^{\alpha} R_{\alpha \nu}\right] \nonumber\\[5pt]
+&\, \xi \left[\frac{1}{2}\nabla_{\alpha}\nabla_{\mu} (B^{\alpha} B_{\nu})+\frac{1}{2}\nabla_{\alpha} \nabla_{\nu} (B^\alpha B_\mu)-\frac{1}{2}\,\square(B_uB_u) - \frac{1}{2}g_{uv}\nabla_\alpha\nabla_\beta(B^\alpha B^\beta)\right]
+\kappa T^M _{\mu v}\,.
\end{align}
Here, $G_{\mu \nu}$ is the Einstein tensor and $T_{\mu\nu}^{M}$ is the energy 
momentum tensor. Further, varying the action \eqref{action} concerning the
bumblebee field gives the equation of motion of the field as 
\cite{Capelo_2015,Maluf_2021}
\begin{equation}\label{BE}
\nabla_{\mu}B^{\mu \nu} = 2\left(V'B^{\nu}- \frac{\xi}{2\kappa}B_{\mu}R^{\mu\nu}\right)\!.
\end{equation}
If the left-hand side of the equation \eqref{BE} vanishes, then the relation 
becomes a simple algebraic relation between the bumblebee potential $V$ and 
the geometry of spacetime. In our work, we have considered the Bianchi type I 
metric and a bumblebee field to solve the equation {\eqref{FE}}. Here, we have 
considered the VEV of the field and hence it holds the condition: $V = V' = 0$. 
%%%%%%%%%%%%%%%%%%%%%%%%%%%%%%%%%%%%%%%%%%%%%%%%%%%%%%%%%
\section{Bianchi Cosmology in bumblebee gravity}\label{3}
We have considered the Bianchi type I metric in our study which has the form:
\begin{equation}\label{metric}
ds^2 = -\,dt^2+a_1^2(t)\, dx^2+ a_2^2(t)\, dy^2+ a_3^2(t)\, dz^2,
\end{equation} 
where $a_1, a_2, a_3$ are the directional scale factors along $x,y,z$ 
directions respectively. Thus, BI metric provides three directional Hubble 
parameters: $H_1 =\dot{a_1}/a_1, H_2 = \dot{a_2}/a_2,$ and 
$H_3 = \dot{a_3}/a_3$ along three axial directions. Accordingly, the 
average expansion scale factor for the metric is $(a_1 a_2 a_3)^{\frac{1}{3}}$ 
and the average Hubble parameter can be written as
\begin{equation}\label{Hub}
H = \frac{1}{3}(H_1+H_2+H_3).
\end{equation}
In our work, we have considered the bumblebee field, which has only one 
surviving component, the temporal component \cite{Capelo_2015}, i.e.
\begin{equation}
B_{\mu}  = \left(B(t),0,0,0\right),
\end{equation}
and it holds the condition, $B_{\mu\nu} = 0$. Thus, under this condition 
equation \eqref{BE} can be simplified as
 \begin{equation}\label{vev}
 \left[V' - \frac{3\xi}{2\kappa}\left(\frac{\ddot{a_1}}{a_1}+\frac{\ddot{a_2}}{a_2}+\frac{\ddot{a_3}}{a_3}\right)\right]=0.
 \end{equation}
Now, for the stress-energy tensor $T_{\mu}^{\nu} = diag(-\rho,P,P,P)$, the 
temporal component of the field equation \eqref{FE} can be written as
 \begin{equation}\label{fe_temp}
 \left(H_1 H_2+H_2 H_3+ H_3 H_1\right)- \xi \left(H_1^2 + H_2^2 + H_3^2\right)B^2- \xi B \dot{B}\left(H_1+H_2+H_3\right) = \kappa\left(\rho+V\right).
 \end{equation}
And, the spatial components of the filed equations are obtained as follows:
\begin{align}\label{fex}
\Big[\big(\frac{\ddot{a_2}}{a_2}+\frac{\ddot{a_3}}{a_3}\big)&+H_2 H_3\Big]\left(1- \xi B^2\right)= \kappa\left(-P+ V\right) - \xi B^2\Big[\frac{1}{2}\big(H_1 + H_2 + H_3\big)^{2} \nonumber\\[5pt]&-\frac{3}{2}\left(H_1^{2} + H_2^{2} + H_3^{2}\right) \Big] + \xi\left[2\left( H_2 + H_3\right)B\dot{B}+B\ddot{B}+ \dot{B}^2\right],
\end{align}
\begin{align}\label{fey}
\Big[\big(\frac{\ddot{a_1}}{a_1}+\frac{\ddot{a_3}}{a_3}\big)&+ H_3 H_1\Big]\left(1- \xi B^2\right)=\kappa \left(-P+ V\right) -\xi B^2\Big[\frac{1}{2}\big(H_1 + H_2 + H_3\big)^{2}\nonumber\\[5pt]&-\frac{3}{2}\left(H_1^{2} + H_2^{2} + H_3^{2}\right)\Big] +\xi\left[2\left(H_1+ H_3\right)B\dot{B}+B\ddot{B}+ \dot{B}^2\right],
\end{align}\begin{align}\label{fez}
\Big[\Big(\frac{\ddot{a_1}}{a_1}+\frac{\ddot{a_2}}{a_2}\big)&+H_1 H_2\Big]\left(1- \xi B^2\right)=\kappa \left(-P+ V\right)-\xi B^2\Big[\frac{1}{2}\big(H_1 + H_2 + H_3\big)^{2}\nonumber\\[5pt]&-\frac{3}{2}\left(H_1^{2} + H_2^{2} + H_3^{2}\right)\Big] +\xi\left[2\left(H_1 + H_2 \right)B\dot{B}+B\ddot{B}+ \dot{B}^2\right].
\end{align}
Further, by adding all the spatial components of the field equations, one can
write,
\begin{align}\label{fe_sp}
\bigg[2\Big(\frac{\ddot{a_1}}{a_1}&+\frac{\ddot{a_2}}{a_2}+\frac{\ddot{a_3}}{a_3}\Big)+\big(H_1 H_2 +H_2 H_3+ H_3 H_1\big)\bigg]\left(1- \xi B^2\right)= 
 3\kappa\left(-P+ V\right)\nonumber\\[5pt]&-3\xi B^2\Big[\frac{1}{2}\left(H_1 + H_2 + H_3\right)^{2} -\frac{3}{2}\left(H_1^{2} + H_2^{2} + H_3^{2}\right) \Big]+\xi\left[4\left(H_1 + H_2 + H_3\right)B\dot{B}+3B\ddot{B}+ 3\dot{B}^2\right].
 \end{align}
These equations are used later for deriving various cosmological parameters. 
Moreover, using the condition ${\nabla_{\mu}T^{\mu\nu} =0}$, we get 
the continuity relation as
 \begin{align}\label{conti_eq}
 \dot{\rho} = &-3H(\rho+p)-\frac{3 \xi B(HB+\dot{B})}{\kappa}\left[\left(\frac{\ddot{a_1}}{a_1} +\frac{\ddot{a_2}}{a_2}+\frac{\ddot{a_3}}{a_3}\right)-2 \sigma^2 \right]\nonumber\\[5pt]&-\frac{3 \xi B^2}{\kappa} \left[H^{3} - 3H \sigma^{2} -H_{1} H_{2} H_{3}- 2 \sigma \dot{\sigma} + \frac{1}{3} \left(\frac{\dddot{a_1}}{a_1} +\frac{\dddot{a_2}}{a_2}+\frac{\dddot{a_3}}{a_3}\right) \right].
 \end{align}
 \section{Field Equations Under Vacuum Expectation Value (VEV) Condition}
\label{4}
At VEV, the bumblebee field equation holds the condition $V =V'=0$, 
{which leads to equation \eqref{vev} as 
${\frac{\ddot{a_1}}{a_1} + \frac{\ddot{a_2}}{a_2}+\frac{\ddot{a_3}}{a_3}=0}$} and also it can be assumed that the bumblebee field is constant for a 
time-like vector and gives the relation 
$B_{\mu} B^{\mu} = \pm\, B_{0}^{2}$ \cite{Neves_2023}. Thus, at VEV, while 
considering {equation \eqref{vev}} along with taking the average 
Hubble parameter expression \eqref{Hub}, the field equations \eqref{fe_temp} 
and \eqref{fe_sp} can be written as
 \begin{align}\label{FE_v1}
 3H^{2} & = \frac{\kappa \rho}{\left(1-l\right)}+\frac{\sigma^2\left(1- 2l\right)}{\left(1-l\right)},\\[5pt]
\label{FE_v2}
 {3H^{2} + 2\dot{H}} & {= -\frac{\kappa P}{\left(1-l\right)} + \frac{3l\sigma^2}{\left(1-l\right)}}.
 \end{align}
Here, $l = \xi B_{0}^{2}$ is the Lorentz violation parameter and 
\begin{equation}\label{sigma}
\sigma^2 = \frac{1}{2}\left[\left(H_1^2+H_2^2+H_3^2\right) - 3H^2\right]
\end{equation} 
is the shear scalar associated with the anisotropic spacetime. Using equations 
(\ref{FE_v1}) and (\ref{FE_v2}) we can obtain the effective equation of state 
and it takes the form:
  \begin{equation}\label{Om_eff}
  \omega_{eff} = -\frac{3H^2+2\dot{H}}{3H^2}={\frac{\kappa P -3l\sigma^2}{\kappa \rho + \sigma^2\left(1- 2l\right)}}.
  \end{equation}
In terms of $\omega_{eff}$, the deceleration parameter $(q)$ can be written as
 \begin{equation}\label{dec}
 q = \frac{1}{2}(1+3\omega_{eff}).
 \end{equation}
The continuity relation \eqref{conti_eq} under the VEV condition takes the 
form:
 
 \begin{equation}\label{cont_vev}
{\dot{\rho} = -3H(\rho+p)-\frac{3l}{\kappa} \left[H^{3}- 3H \sigma^{2} -H_{1} H_{2} H_{3}- 2 \sigma \dot{\sigma} + \frac{1}{3} \left(\frac{\dddot{a_1}}{a_1} +\frac{\dddot{a_2}}{a_2}+\frac{\dddot{a_3}}{a_3}\right) \right]+\frac{6l H \sigma^2}{\kappa}.}
 \end{equation}
Now, taking the $\sigma^2 \propto \theta^2$ condition \cite{Sarmah_2023,Sarmah_2024} in which the 
$\theta$ is the expansion scalar, we can write, $H_1 = \alpha H_2$ and 
$H_1 = \beta H_3$, where $\alpha$ and $\beta$ are two proportionality 
constants. {This assumption of proportionality for the homogeneous 
anisotropic background has been widely discussed in several literature 
(e.g.~see \cite{Colins_1971, Colins_1980, Ban_1985, Ban_1985, Rib_1987}), and 
is based on the fact that as the expansion increases with different Hubble 
rates in different directions, the associated shear scalar increases in some 
proportion (see equation (16))}.These lead to the average Hubble parameter having a simplified form
in terms of the x-directional parameter $H_1$ as
 \begin{equation}
 H = \frac{\alpha + \beta + \alpha \beta}{3 \alpha \beta} H_1 = \frac{1}{\lambda} H_1
 \end{equation}
and also lead to the shear scalar in form as
 \begin{equation}
 \sigma^2 = \frac{3H^2 \left[3(\alpha^2+\beta^2+\alpha \beta )- \alpha \beta (\alpha+\beta + \alpha \beta)^2\right]}{(\alpha+\beta+\alpha\beta)}=3H^2 \eta,
 \end{equation}
where $\lambda = (\alpha + \beta + \alpha \beta)/3 \alpha \beta$ and 
$\eta = \left[3(\alpha^2+\beta^2+\alpha \beta )- \alpha \beta (\alpha+\beta + \alpha \beta)^2\right]/(\alpha+\beta+\alpha\beta)$. In view of this form of
$\sigma^2$, equation \eqref{cont_vev} takes the form:
  \begin{align}
 {\dot{\rho} =} & {-3H\rho\left[\frac{2-l(1+3\eta) - l(1-3\eta)}{(1-l)}\right] -\frac{3l H^3}{\kappa} \left[1 -\frac{27\alpha ^2 \beta^2}{(\alpha + \beta + \alpha \beta)^3}  \frac{18\eta^2(1-2l)}{(1-l)}+9(\gamma - \eta) \right]}
\nonumber\\[8pt]&{+\frac{45lH^3 \eta}{\kappa}},
 \end{align}
where $\gamma = 3(\alpha \beta\lambda + \eta)$. With the consideration of 
$p = \omega \rho$ along with equations \eqref{FE_v1} and \eqref{FE_v2}, the 
above equation can further be rewritten as
 
 \begin{align}\label{cont}
 {\dot{\rho} =} & {-3H\rho\Bigg[\frac{(1+\omega)-l\left\{1+\eta+\omega(1-\eta)\right\}}{(1-l)}-l\,\frac{1 -\frac{\lambda^3}{\alpha \beta} + \frac{18\eta^2(1-2l)}{(1-l)}+9(\gamma - \eta)}{(3+\eta) - l(3+2\eta)}}\nonumber\\[5pt] &{-\frac{15l \eta}{ \left\{(3+\eta) - l(3+2\eta)\right\}}\Bigg].}
 \end{align}
 
This equation of continuity can be solved for the density $\rho$ as
 \begin{equation}
 \rho = \rho_0 a^{-3 \delta},
 \end{equation}
where
 
 \begin{align}
 {\delta =} &\,{\frac{(1+\omega)-l\left\{1+\eta+\omega(1-\eta)\right\}}{(1-l)}-l\,\frac{1 -\frac{\lambda^3}{\alpha \beta} + \frac{18\eta^2(1-2l)}{(1-l)}+9(\gamma - \eta)}{(3+\eta) - l(3+2\eta)}} \nonumber\\[5pt]&{-\frac{15l \eta}{ \left\{(3+\eta) - l(3+2\eta)\right\}}\Bigg].}
 \end{align}
Furthermore, from the equation \eqref{FE_v1} we can write the Hubble parameter 
as
 \begin{equation}
 {H = \sqrt{\frac{\kappa \rho}{{3\left\{(1-l)-\eta(1-2l)\right\}}}}\,}.
 \end{equation}
Again, taking the relation $a = \frac{1}{(1+z)}$ in which $z$ is the 
cosmological redshift, we can rewrite the Hubble parameter in the form:
 \begin{equation}\label{hub}
 H = H_0 \sqrt{E(z)},
 \end{equation}
where
 \begin{equation}\label{ez}
{E(z)=\frac{\left[\Omega_{mo}(1+z)^{3\delta_m}+\Omega_{ro}(1+z)^{3\delta_r}+\Omega_{\Lambda 0}(1+z)^{3\delta_\Lambda} \right]}{\left\{(1-l)-\eta(1-2l)\right\}}},
 \end{equation}
with
 \begin{align}\label{dm}
 {\delta_m = }&\;{\frac{1-l\left(1+\eta\right)}{(1-l)}-l\,\frac{1 -\frac{\lambda^3}{\alpha \beta} + \frac{18\eta^2(1-2l)}{(1-l)}+9(\gamma - \eta)}{(3+\eta) - l(3+2\eta)}}\nonumber\\[4pt] &{-\frac{15l \eta}{ \left\{(3+\eta) - l(3+2\eta)\right\}}},
\end{align}
 \begin{align}\label{dr}
 {\delta_r = }&\;{\frac{4-2l\left(2+\eta\right)}{3(1-l)}-l\,\frac{1 -\frac{\lambda^3}{\alpha \beta} + \frac{18\eta^2(1-2l)}{(1-l)}+9(\gamma - \eta)}{(3+\eta) - l(3+2\eta)}}\nonumber\\[4pt]&{-\frac{15l \eta}{ \left\{(3+\eta) - l(3+2\eta)\right\}}},
 \end{align}

 \begin{align}\label{dl}
{\delta_\Lambda =} &\;{-\frac{2l \eta}{(1-l)}-l\,\frac{1 -\frac{\lambda^3}{\alpha \beta} + \frac{18\eta^2(1-2l)}{(1-l)}+9(\gamma - \eta)}{(3+\eta) - l(3+2\eta)}}\nonumber\\[4pt]&{-\frac{15l \eta}{ \left\{(3+\eta) - l(3+2\eta)\right\}}.}
 \end{align}
 {Further, for $\boldsymbol{z=0}$, ${E(0) = 1}$ and thus 
equation \eqref{ez} gives ${\eta = \frac{l}{2l-1}}$  and this 
relation has been applied to all cosmological parameters while constraining 
the parameters and other analyses.} Utilizing the expression of $H$, 
i.e.~$E(z)$, the distance modulus $(D_m)$ can be obtained from the following 
formula:
 \begin{equation}\label{Dm}
 D_m = 5\log d_L+ 25,
 \end{equation}
where $d_L$ is the luminosity distance and it has the mathematical form: 
 \begin{equation}\label{d_L}
 d_L = \frac{(1+z)}{H_0}{\int_{0}^{\infty} \frac{dz}{\sqrt{E(z)}}}.
 \end{equation}
Also, the equation \eqref{Om_eff} for the effective equation of state can be 
rewritten as
 \begin{equation}\label{omeff}
 {\omega_{eff} = \frac{\frac{1}{3}\,\Omega_{r0}(1+z)^{3\delta_r}-\Omega_{\Lambda0}(1+z)^{\delta_{\Lambda}}-\frac{9l^2}{2l-1}\frac{H^2}{H_0^2}}{\Omega_{m0}(1+z)^{\delta_{m}}+\Omega_{r0}(1+z)^{3\delta_r}+\Omega_{\Lambda0}(1+z)^{\delta_{\Lambda}}-l}\,.}
 \end{equation}
 
With all these derivations, we are ready for the graphical visualization of 
all the cosmological parameters that have been expressed in equations 
\eqref{dec}, \eqref{hub}, \eqref{Dm} and \eqref{omeff}. However, before doing 
that, we need to constrain different model parameters, like $l,\alpha, \beta, 
\lambda$, etc.~to obtain results consistent with the current observations 
which we have done in the next section.
\section{Parameters estimations and constraining}\label{5}
For estimating and constraining the parameters, we have used the Bayesian 
inference technique which is based on Bayes' theorem. The theorem states that 
the posterior distribution $\mathcal{P}({\psi}|\mathcal{D},\mathcal{M})$ of 
the parameter $\psi$ for the model $\mathcal{M}$ with cosmological data 
$\mathcal{D}$ can be obtained as
\begin{equation}
 \mathcal{P}({\psi}|\mathcal{D},\mathcal{M}) = \frac{\mathcal{L}(\mathcal{D}|\psi, \mathcal{M}) \pi (\psi|\mathcal{M})}{\mathcal{E}(\mathcal{D}|\mathcal{M})}.
\end{equation}
Here, $\mathcal{L}(\mathcal{D}|\psi, \mathcal{M})$, $\pi (\psi|\mathcal{M})$ 
and $\mathcal{E}(\mathcal{D}|\mathcal{M})$ are the likelihood, the prior 
probability of the model parameters and the Bayesian evidence respectively. 
The Bayesian evidence can be obtained as
\begin{equation}
\mathcal{E}(\mathcal{D}|\mathcal{M}) = \int_{\mathcal{M}} \mathcal{L}(\mathcal{D}|\psi, \mathcal{M}) \pi (\psi|\mathcal{M}) d\psi,
\end{equation}
in which the likelihood $\mathcal{L}(\mathcal{D}|\psi, \mathcal{M})$ can be 
considered as a multivariate Gaussian likelihood function and it takes the 
form \cite{akarsu_2019}:
\begin{equation}
\mathcal{L}(\mathcal{D}|\psi, \mathcal{M}) \propto \exp\left[\frac{-\chi^2(\mathcal{D}|\psi, \mathcal{M})}{2}\right],
\end{equation}
where $\chi^2(\mathcal{D}|\psi, \mathcal{M})$ is the Chi-squared function of 
the dataset $\mathcal{D}$. For  a uniform prior distribution 
$\pi(\psi|\mathcal{M})$ of the model parameters, we can simply be considered 
the posterior distribution as
\begin{equation}
\mathcal{P}({\psi}|\mathcal{D},\mathcal{M}) \propto \exp\left[\frac{-\chi^2(\mathcal{D}|\psi, \mathcal{M})}{2}\right].
\end{equation}  
We have used this technique to estimate various model and cosmological 
parameters with the help of various observational data sets in our current 
work.
\subsection{Data and their respective likelihoods}
Here we will use observational data of Hubble parameter, BAO, CMB, and 
Pantheon supernovae type Ia from various sources to carry forward our analysis.
\subsubsection{\textbf{Hubble parameter $H(z)$}}
We have considered $57$ observational $H(z)$ data from different literatures 
and tabulated them in Table \ref{table2}. The chi-square function value for 
the mentioned dataset of $H(z)$ denoted by $\chi^2$ can be calculated as  
\begin{equation}
\chi^2_{H} =  \sum_{n=1}^{57} \frac{\left[H^{obs}(z_n)- H^{th}(z_n)\right]^2}{\sigma^2_{H^{obs}(z_n)}},
\end{equation}
where $\sigma^2_{H^{obs}(z_n)}$ is the standard deviation of the nth 
observational $H(z)$ data and $H^{th}(z_n)$ is the theoretical value of $H$ 
obtained from the considered cosmological model at $z_n$.
%\vspace{-5mm}
\begin{center}
\begin{table}[!hbt]
\caption{Available observational Hubble parameter ($H^{obs}(z)$) data set in 
the unit of km/s/Mpc extracted from different literatures.}
\vspace{2mm}
\begin{tabular}{ccc|ccc}
\hline 
\rule[-0.2ex]{0pt}{2.5ex} \hspace{0.5cm} $z$ \hspace{0.5cm}  & \hspace{0.2cm} $ H^{obs}(z)$ \hspace{0.0cm} & \hspace{0.0cm} Reference \hspace{0.0cm} &
\hspace{0.5cm} $z$ \hspace{0.5cm}  & \hspace{0.2cm} $ H^{obs}(z)$ \hspace{0.0cm} & \hspace{0.0cm} Reference \hspace{0.0cm}\\ 
\hline\\[-9pt]
\rule[-1ex]{0pt}{2.5ex} 0.0708 & 69.0 $\pm$ 19.68 & \cite{Zhang_2014} & 
0.48 & 97.0 $\pm$ 62.0 & \cite{Ratsimbazafy_2017}\\
\rule[-1ex]{0pt}{2.5ex} 0.09 & 69.0 $\pm$ 12.0 & \cite{Simon_2005} &
0.51 & 90.8 $\pm$ 1.9 & \cite{Alam_2017}\\
\rule[-1ex]{0pt}{2.5ex} 0.12 & 68.6 $\pm$ 26.2 & \cite{Zhang_2014} &
0.52 & 94.35 $\pm$ 2.64 & \cite{Wang_2017}\\
%0.57 & 92.4 $\pm$ 4.5 & \cite{Samushia_2013} \\
\rule[-1ex]{0pt}{2.5ex} 0.17 & 83.0 $\pm$ 8.0 & \cite{Simon_2005} &
0.56 & 93.34 $\pm$ 2.3 & \cite{Wang_2017}\\
%0.593 & 104.0 $\pm$ 13.0 & \cite{Moresco_2012}\\
\rule[-1ex]{0pt}{2.5ex}0.179 & 75.0 $\pm$ 4.0 & \cite{Moresco_2012} &
0.57 & 92.4 $\pm$ 4.5 & \cite{Samushia_2013} \\
%0.60 & 87.9 $\pm$ 6.1 & \cite{Blake_2012}\\
\rule[-1ex]{0pt}{2.5ex} 0.199 & 75.0 $\pm$ 5.0 & \cite{Moresco_2012} &
0.57 & 87.6 $\pm$ 7.8 & \cite{Chuang_2013} \\
%0.61 & 97.8 $\pm$ 2.1 & \cite{Alam_2017}\\
\rule[-1ex]{0pt}{2.5ex} 0.20 & 72.9 $\pm$ 29.6 & \cite{Zhang_2014} &
0.59 & 98.48 $\pm$ 3.18 & \cite{Wang_2017}\\
%0.68 & 92.0 $\pm$ 8.0 & \cite{Moresco_2012}\\
\rule[-1ex]{0pt}{2.5ex} 0.24 & 79.69 $\pm$ 2.65 & \cite{Gaztanaga_2009} &
0.593 & 104.0 $\pm$ 13.0 & \cite{Moresco_2012}\\
%0.73 & 97.3 $\pm$ 7.0 & \cite{Blake_2012}\\
\rule[-1ex]{0pt}{2.5ex} 0.27 & 77.0 $\pm$ 14.0 & \cite{Simon_2005} &
0.60 & 87.9 $\pm$ 6.1 & \cite{Blake_2012}\\
%0.781 & 105.0 $\pm$ 12.0 & \cite{Moresco_2012}\\
\rule[-1ex]{0pt}{2.5ex} 0.28 & 88.8 $\pm$ 36.6 & \cite{Zhang_2014} &
0.61 & 97.8 $\pm$ 2.1 & \cite{Alam_2017}\\
%0.875 & 125.0 $\pm$ 17.0 & \cite{Moresco_2012}\\
\rule[-1ex]{0pt}{2.5ex} 0.30 & 81.7 $\pm$ 6.22  & \cite{Oka_2014} &
0.64 & 98.82 $\pm$ 2.98 & \cite{Wang_2017}\\
%0.88 & 90.0 $\pm$ 40.0 & \cite{Ratsimbazafy_2017}\\
\rule[-1ex]{0pt}{2.5ex} 0.31 & 78.18 $\pm$ 4.74  & \cite{Wang_2017} &
0.6797 & 92.0 $\pm$ 8.0 & \cite{Moresco_2012}\\
%0.88 & 90.0 $\pm$ 40.0 & \cite{Ratsimbazafy_2017}\\
\rule[-1ex]{0pt}{2.5ex} 0.34 & 83.8 $\pm$ 3.66  & \cite{Gaztanaga_2009} &
0.73 & 97.3 $\pm$ 7.0 & \cite{Ratsimbazafy_2017}\\
\rule[-1ex]{0pt}{2.5ex} 0.35 & 82.7 $\pm$ 9.1  & \cite{Xu_2013} &
0.781 & 105.0 $\pm$ 12.0 & \cite{Moresco_2012}\\
\rule[-1ex]{0pt}{2.5ex} 0.352 & 83.0 $\pm$ 14.0 & \cite{Moresco_2012}&
0.8754 & 125.0 $\pm$ 17.0 & \cite{Moresco_2012}\\
%0.90 & 117.0 $\pm$ 23.0 & \cite{Simon_2005}\\
\rule[-1ex]{0pt}{2.5ex} 0.36 & 79.94 $\pm$ 3.38  & \cite{Wang_2017} &
0.88 & 90.0 $\pm$ 40.0 & \cite{Ratsimbazafy_2017}\\
\rule[-1ex]{0pt}{2.5ex} 0.38 & 81.9 $\pm$ 1.9 & \cite{Alam_2017} &
0.90 & 117.0 $\pm$ 23.0 & \cite{Simon_2005}\\
%1.037 & 154.0 $\pm$ 20.0 & \cite{Moresco_2012}\\
\rule[-1ex]{0pt}{2.5ex} 0.3802 & 83.0 $\pm$ 13.5 & \cite{Moresco_2016} &
1.037 & 154.0 $\pm$ 20.0 & \cite{Moresco_2012}\\
%1.30 & 168.0 $\pm$ 17.0 & \cite{Simon_2005}\\
\rule[-1ex]{0pt}{2.5ex} 0.40 & 82.04 $\pm$ 2.03 & \cite{Wang_2017} &
1.30 & 168.0 $\pm$ 17.0 & \cite{Simon_2005}\\
%1.363 & 160.0 $\pm$ 33.6 & \cite{Moresco_2015}\\
\rule[-1ex]{0pt}{2.5ex} 0.40 & 95.0 $\pm$ 17.0 & \cite{Simon_2005} &
1.363 & 160.0 $\pm$ 33.6 & \cite{Moresco_2015}\\
\rule[-1ex]{0pt}{2.5ex} 0.4004 & 77.0 $\pm$ 10.2 & \cite{Moresco_2016} &
1.43 & 177.0 $\pm$ 18.0 & \cite{Simon_2005}\\
\rule[-1ex]{0pt}{2.5ex} 0.4247 & 87.1 $\pm$ 11.2 & \cite{Moresco_2016} &
1.53 & 140.0 $\pm$ 14.0 & \cite{Simon_2005}\\
\rule[-1ex]{0pt}{2.5ex} 0.43 & 86.45 $\pm$ 3.68 & \cite{Gaztanaga_2009} &
1.75 & 202.0 $\pm$ 40.0 & \cite{Simon_2005}\\
\rule[-1ex]{0pt}{2.5ex} 0.44 & 82.6 $\pm$ 7.8 & \cite{Blake_2012} &
1.965 & 186.5 $\pm$ 50.4 & \cite{Moresco_2015}\\
\rule[-1ex]{0pt}{2.5ex} 0.44 & 84.81 $\pm$ 1.83 & \cite{Wang_2017} &
2.30 & 224 $\pm$ 8.6 & \cite{Busca_2013}\\
\rule[-1ex]{0pt}{2.5ex} 0.4497 & 92.8 $\pm$ 12.9 & \cite{Moresco_2016} & 
2.33 & 224 $\pm$ 8 & \cite{Bautista_2017}\\
\rule[-1ex]{0pt}{2.5ex} 0.47 & 89.0 $\pm$ 50.0 & \cite{Ratsimbazafy_2017} &
2.34 & 223.0 $\pm$ 7.0 & \cite{Delubac_2015}\\
\rule[-1ex]{0pt}{2.5ex} 0.4783 & 80.9 $\pm$ 9.0 & \cite{Moresco_2016} &
2.36 & 227.0 $\pm$ 8.0 & \cite{Ribera_2014}\\
\rule[-1ex]{0pt}{2.5ex} 0.48 & 87.79 $\pm$ 2.03 & \cite{Wang_2017} &&&\\
\hline
\end{tabular}
\label{table2}
\end{table}
\end{center}
\vspace{-30pt}
\subsubsection{\textbf{BAO measurements}}
Baryon acoustic oscillation (BAO) helps to understand the angular diameter 
distance between two points in the Universe in terms of redshift ($z$) 
and is also useful in studying the evolution of $H(z)$. Usually, the BAO 
measurements provide the dimensionless ratio $d$ of the comoving size of the 
sound horizon $r_s$ at the drag redshift $z_d = 1059.6$ \cite{Planck_2015} to 
the volume-averaged distance $D_v (z)$. Thus,
\begin{equation}
d = \frac{r_s(z_d)}{D_v (z)},
\end{equation}
where $r_s(z_d)$ and $D_v (z)$ are expreesed respectively as
\begin{equation}\label{r}
r_s(z_d) = \int^{\infty}_{z_d} \frac{c_s dz}{H(z)},
\end{equation}
\begin{equation}\label{DV}
D_v (z) = \left[(1+z)^2 D_A (z)^2 \frac{cz}{H(z)}\right]^{1/3}\!\!\!\!\!.
\end{equation}
The term $C_s = c/\!\sqrt{3(1+\mathcal{R})}$ appears in equation \eqref{r} is
the sound velocity in the baryon-photon fluid. Here, $\mathcal{R} = 3 \Omega_{b0}/(4\Omega_{r0}(1+z))$ with $\Omega_{b0} = 0.022\,h^{-2}$ \cite{Cooke_2016}, 
$\Omega_{r0} = \Omega_{\gamma 0}\left(1+ 7/8\,(4/11)^{4/3}N_eff\right)$ in
which $\Omega_{\gamma 0} = 2.469\times 10^{-5} h^{-2}$ and $N_{eff} = 3.046$
\cite{Dodelson_2003, akarsu_2019}. Moreover, $D_A$ in equation \eqref{DV} is 
the angular diameter distance which can be expressed as
\begin{equation}
D_A = \frac{c}{(1+z)} \int^{z}_{0}\frac{dz}{H(z)},
\end{equation}
and $c$ is the speed of light. In our work, we have tabulated $8$ BAO data in 
Table \ref{table3} from various works in the literature and computed the total 
chi-square value ($\chi_{BAO}^2$) for them.

The chi-square function value $\chi^2_{d}$ of first five data set of Table 
\ref{table3} can be calculated by using the expression,
\begin{equation}
\chi^2_{d} = \sum_{i=1}^{5} \frac{\left[d^{obs}(z_i)- d^{th}(z_i)\right]^2}{\sigma^2_{d^{obs}(z_i)}},
\end{equation}
in which $d^{th}(z_i)$ is the theoritical value of $d$ for the considered 
cosmological model. For the remaining three dataset of the WiggleZ survey 
in Table \ref{table3}, the chi-square value can be obtained by using the 
method of covariant metrix. Here the inverse covariant matrix of the 
considered dataset can be written as \cite{akarsu_2019}
\begin{center}
\begin{table}[t]
\caption{Available observational BAO data.}
\vspace{8pt}
\begin{tabular}{ccccc}
\hline
\rule[-0.2ex]{0pt}{2.5ex} \hspace{0.0cm} Survey \hspace{0.5cm}&\hspace{0.5cm} $z_i$ \hspace{0.5cm}  & \hspace{0.2cm} $ d^{obs}(z_i)$ \hspace{0.2cm}&\hspace{0.2cm}$\sigma_{d^{obs}(z_i)}$\hspace{0.0cm} & \hspace{0.0cm} Reference \hspace{0.0cm}
\\ 
\hline\\[-9pt]
\rule[-1ex]{0pt}{2.5ex} 6dFGS & 0.106 & 0.3360 & 0.0150 & \cite{Beutler_2011} \\
\rule[-1ex]{0pt}{2.5ex} MGS & 0.15 & 0.2239 & 0.0084 & \cite{Ross_2015} \\
\rule[-1ex]{0pt}{2.5ex} BOSS LOWZ & 0.32 & 0.1181 & 0.0024 & \cite{Padmanabhan_2012} \\
\rule[-1ex]{0pt}{2.5ex} SDSS(R) & 0.35 & 0.1126 & 0.0022 & \cite{Anderson_2014} \\
\rule[-1ex]{0pt}{2.5ex} BOSS CMASS & 0.57 & 0.0726 & 0.0007 & \cite{Padmanabhan_2012} \\
\rule[-1ex]{0pt}{2.5ex} WiggleZ & 0.44 & 0.073 & 0.0012 & \cite{Blake_2012} \\
\rule[-1ex]{0pt}{2.5ex} WiggleZ & 0.6 & 0.0726 & 0.0004 & \cite{Blake_2012} \\
\rule[-1ex]{0pt}{2.5ex} WiggleZ & 0.73 & 0.0592 & 0.0004 & \cite{Blake_2012} \\
\hline
\end{tabular}
\label{table3}
\end{table}
\end{center}
\vspace{-20pt}
\begin{equation}
C^{-1}_{w}=
\begin{bmatrix}
  1040.3 & -807.5 & 336.8 \\
  -807.5 & 3720.3 & -1551.9\\
  336.8  & -1551.9 & 2914.9
\end{bmatrix}.
\end{equation}
From this inverse covariant matrix the chi-square value for the considered 
three WiggleZ survey dataset can be obtained as
\begin{equation}
\chi^{2}_{w} = D^{T}C^{-1}_wD,
\end{equation}
in which the matrix $D$ can be written as
\begin{equation}
D = 
\begin{bmatrix}
d^{obs}(0.44)- d^{th}(0.44)\\
d^{obs}(0.60)- d^{th}(0.60)\\
d^{obs}(0.73)- d^{th}(0.73)\\
\end{bmatrix}.
\end{equation}
Thus, the total chi-squar value $\chi^{2}_{BAO}$ can be obtained as
\begin{equation}
\chi^{2}_{BAO} = \chi^{2}_{d} + \chi^{2}_{w}.
\end{equation}
\subsubsection{\textbf{CMB data}}
The CMB data include the angular scale of the sound horizon at the last 
scattering surface which is denoted by $l_a$ and can be defined as
\begin{equation}
l_a = \pi \frac{r(z^*)}{r_s (z^*)},
\end{equation}
where $r(z_*)$ is the comoving distance to the last scattering surface and it 
can be measured as
\begin{equation}
r(z_*) = \int^{z^*}_{0} \frac{cdz}{H(z)},
\end{equation}
and $r_s(z_*)$ is the size of the comoving sound horizon at the redshift of 
the last scattering ($z^* = 1089.9$). The observed value $l^{obs}_{a} = 301.63$ with uncertainty $\sigma_{l_a} = 0.15$ as per Ref.~\cite{Planck_2015}.
The chi-square value $\chi^2_{CMB}$ here can be evaluated as
\begin{equation}
\chi^2_{CMB} = \frac{\left[l^{obs}_a - l^{th}_a\right]^2}{\sigma^2_{l_a}}.
\end{equation}
Here, the $l^{th}_a$ is the theoretical value of $l_a$ for the considered 
cosmological model.
\subsubsection{\textbf{Pantheon plus supernovae type Ia data}}
The Pantheon data sample comprised 1048 observational data spanning over the 
$z$ range between $0.001 < z < 2.3$ taken from five subsample, which include 
PS1, SDSS, SNLS, low-$z$ and HST \cite{Betoule_2014}. The Pantheon plus sample
 is the successor of the Pantheon sample and contains 1701 observational data 
from 18 different sources \cite{Scolnic_2022}. This pantheon plus data compilation 
contains mainly the observed peak magnitude $m_B$ and also the distance 
modulus $D_m$ for different Type Ia supernovae (SN Ia).

Theoretically, the distance modulus $D_m$ can be calculated as
\begin{equation}
D_m = 5 \log_{10} \frac{d_L (z_{hel},z_{cmb})}{10pc} = 5 \log_{10} \frac{d_L (z_{hel},z_{cmb})}{1Mpc} +25,
\end{equation}
in which $z_{hel}$ is the heliocentric redshift, $z_{cmb}$ is the redshift 
of the CMB rest frame and $d_L$ is the luminosity distance. The theoretical
luminosity distance can be calculated by using equation \eqref{d_L}.

The chi-square value for the Pantheon plus dataset is denoted by 
$\chi^{2}_{Pan+}$ can be obtained by using the equation as follows:
\begin{equation}
\chi^{2}_{Pan+} = m^{T}C^{-1}m,
\end{equation}
where $C$ is the total covariance matrix of $m_B$ and the matrix $m$ is 
obtained by the relation $m = m_B - m_{th}$ with 
\begin{equation}
m_{th} = 5 \log_{10} D_{L} + M.
\end{equation}
Here,
\begin{equation}
D_{L} = (1+z_{hel}) \int^{Z_{cmb}}_{0} \frac{H_0 dz}{H(z)},
\end{equation}
and $M$ is the nuisance parameter whose value for the Pantheon dataset is
 $23.739^{+0.140}_{-0.102}$ \cite{Zhao_2019}. Further, the total covariance 
metrix $C$ consists of systematic covariance metrix $C_{sys}$ and the diagonal 
covariance matrix of the statistical uncertainty $D_{stat}$ \cite{Scolnic_2018, akarsu_2019}.
\subsection{Constraining the cosmological parameters}
In order to obtain observational constraints on the anisotropic cosmological 
model in bumblebee gravity theory, we have considered a multivariate joint 
Gaussian likelihood of the form \cite{akarsu_2019}:
\begin{equation}
\mathcal{L}_{tot} \propto \exp\left(\frac{-\chi^2_{tot}}{2}\right),
\end{equation}
in which, 
\begin{equation}
\chi^2_{tot} = \chi^2_{H} + \chi^2_{BAO} +\chi^2_{CMB}+\chi^2_{Pan+}
\end{equation}
In this work, we have considered uniform prior distribution for all 
cosmological parameters and model parameters. The prior range of various 
parameters has been considered as follows: $55 < H_0 < 85$, 
$0.1 < \Omega_{mo} < 0.5$, $0.00001 < \Omega_{ro}  < 0.0001$, 
$0.6 < \Omega_{\Lambda 0}< 1$, $0.9 <\delta_{m} < 1.05$, 
$1.25 <\delta_{r}< 1.75$, $0.001<\delta_{l}< 0.01$, ${0.01 <l< 0.1}$. 
{It is to be noted that we have considered $\delta_m,~\delta_r,~\delta_{\Lambda}$ as the free parameters along with other cosmological 
parameters $\Omega_{m0}$, $\Omega_{r0}$, $\Omega_{\Lambda0}$ and $l$ due to 
complexity of the relations in equation \eqref{dm}, \eqref{dr} and \eqref{dl} 
respectively.} Here, the likelihoods are considered within the mentioned 
ranges such that results should be consistent with standard Planck data release 
2018. However, the parameter $\eta$ is associated with the anisotropic 
characteristics of the Universe hence its current value is considered too 
small. Further, the Lorentz violation parameter $l = \xi B^2_0$ is in general 
considered to be in the order of $10^{-23}$ \cite{Paramos_2014}, however, to 
observe its effect on the cosmological parameters like the Hubble parameter 
and Distance modulus we have taken its range as {${0.01}$} to {${0.1}$}.
The idea of taking higher values of Lorentz violation parameter $l$ 
in the bumblebee theory of gravity for studying other physical systems like 
black holes can be found in Refs.~\cite{Gogoi_2022,Karmakar_2023}. This 
actually helps to study the physical system to understand the effect of 
Lorentz symmetry breaking on it.
\begin{figure}[!h]
\centerline{
  \includegraphics[scale = 0.57]{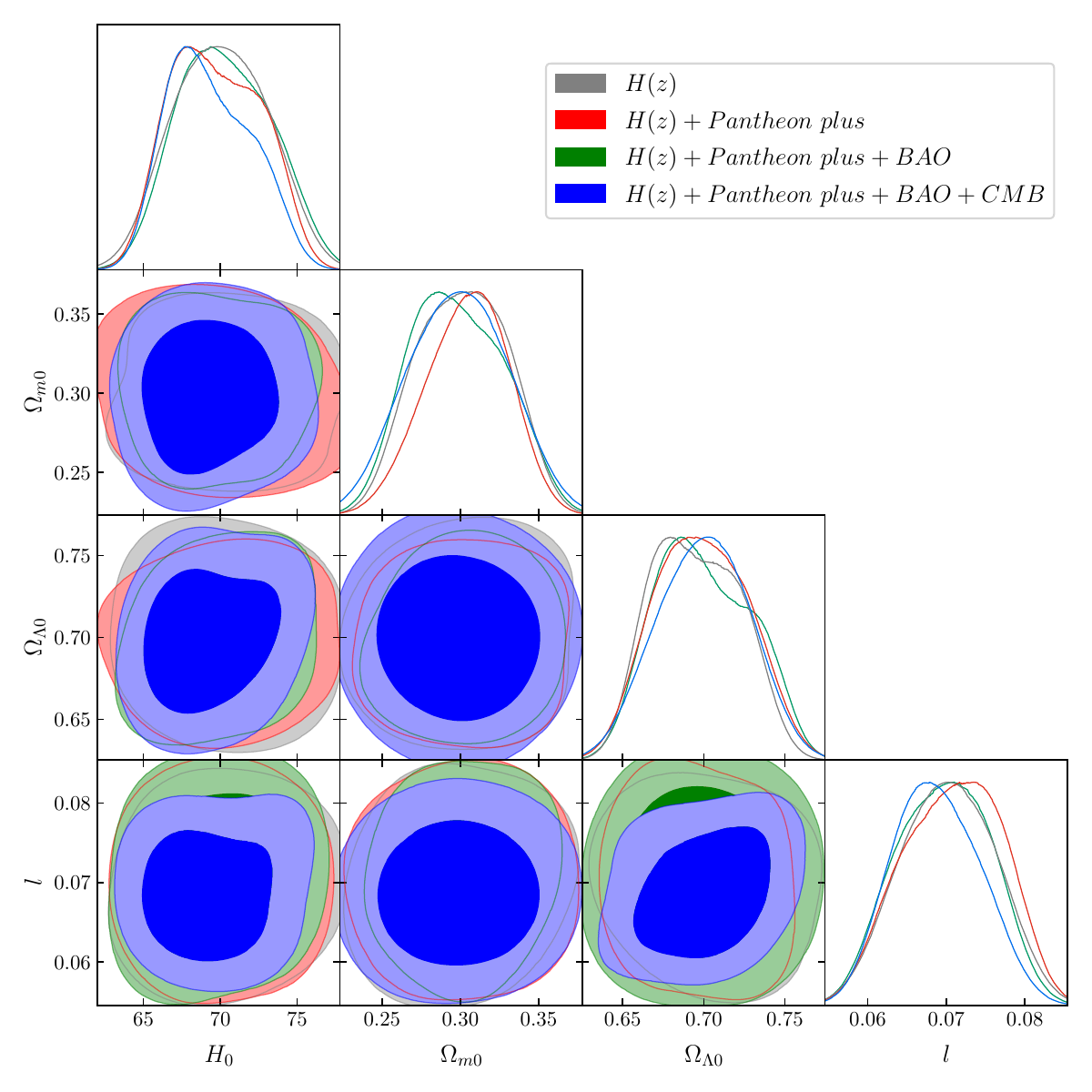}\hspace{0.25cm}
 }
\vspace{-0.2cm}
\caption{One dimensional and two-dimensional marginalized confidence regions 
($68\% ~\text{and} ~95\%$ confidence level) for different anisotropic 
bumblebee model parameters obtained with the help of $H(z)$, Pantheon plus, 
BAO and CMB data.}
\label{fig1}
\end{figure}
\begin{center}
\begin{table}[!h]
\caption{Constrained model parameters' values for the both anisotropic 
bumblebee model and the $\Lambda$CDM model as obtained through confidence 
level corner plots using different cosmological data sources.}
\vspace{8pt}
\scalebox{0.87}{
\begin{tabular}{cccccc}
\hline 
\rule[0.1ex]{0pt}{2.5ex} Model \hspace{2pt} & \hspace{2pt} Parameters \hspace{2pt} & \hspace{2pt} $H(z)$ \hspace{2pt} & \hspace{2pt} $H(z) + \text{Pantheon plus}$ \hspace{2pt} & \hspace{2pt} $H(z) + \text{Pantheon plus} + \text{BAO}$ \hspace{2pt} &\hspace{2pt} $H(z) + \text{Pantheon plus} + \text{BAO} + \text{CMB}$ \hspace{0.25cm}\\
\vspace{-9.5pt}\\ 
\hline\\[-8pt]
%\rule[-2ex]{0pt}{2.5ex}Anisotropic bumblebee&&  &  &&\\
\rule[-2ex]{0pt}{2.5ex}&$H_0$& $\mathbf{70.095^{+3.452}_{-3.533}}$ & ${69.960^{+4.240}_{-2.701}}$& ${69.578^{+3.534}_{-3.173}}$&${69.474^{+4.358}_{-3.167}}$\\
\rule[-2ex]{0pt}{2.5ex}&$\Omega_{m0}$ &${0.301^{+0.031}_{-0.034}}$ & ${0.291^{+0.044}_{-0.026}}$ &${0.298^{+0.037}_{-0.033}}$&${0.297^{+0.038}_{-0.029}}$\\
\rule[-2ex]{0pt}{2.5ex}&$\Omega_{r0}$ &${0.000042^{+0.000015}_{-0.000016}}$ & ${0.000044^{+0.000014}_{-0.000012}}$ &${0.000038^{+0.000013}_{-0.000011}}$& ${0.000040^{+0.000015}_{-0.000013}}$\\
\rule[-2ex]{0pt}{2.5ex}&$\Omega_{\Lambda0}$&${0.702^{+0.031}_{-0.035}}$&$ {0.701^{+0.033}_{-0.029}}$&${0.702^{+0.027}_{-0.034}}$&${0.698^{+0.036}_{-0.031}}$\\
\rule[-2ex]{0pt}{2.5ex} AB$^*$ &$\delta_m$&${0.983^{+0.031}_{-0.020}}$ & ${0.981^{+0.025}_{-0.023}}$ &${ 0.984^{+0.025}_{-0.024}}$&${0.976^{+0.031}_{-0.022}}$\\
\rule[-2ex]{0pt}{2.5ex}&$\delta_{r}$ &${1.305^{+0.016}_{-0.024}}$ &${ 1.304^{+0.014}_{-0.017}}$ &${1.306^{+0.016}_{-0.018}}$&${1.307^{+0.019}_{-0.018}}$ \\
\rule[-2ex]{0pt}{2.5ex}&$\delta_{l}$ &${0.007^{+0.001}_{-0.001}}$ & ${0.007^{+0.001}_{-0.003}}$ &${0.007^{+0.001}_{-0.001}}$&${0.007^{+0.002}_{-0.001}}$ \\
\rule[-2ex]{0pt}{2.5ex}&$l$ &${0.070^{+0.007}_{-0.007}}$&${0.070^{+0.004}_{-0.008}}$ &${0.069^{+0.006}_{-0.008}}$&${0.070^{+0.003}_{-0.006}}$\\
%\rule[-2ex]{0pt}{2.5ex}&$\eta$ &$0.030^{+0.006}_{-0.006}$ &$0.029^{+0.006}_{-0.007}$&$0.030^{+0.005}_{-0.008}$&$0.032^{+0.006}_{-0.008}$\\
\hline\\[-8pt]
%\rule[-2ex]{0pt}{2.5ex}$\Lambda$CDM& & &  &&\\
\rule[-2ex]{0pt}{2.5ex}&$H_0$&$70.167^{+3.192}_{-2.823}$ & $69.804^{+3.841}_{-3.169}$&$69.202^{+3.893}_{-2.833}$&$68.826^{+3.857}_{-2.620}$\\
\rule[-2ex]{0pt}{2.5ex}$\Lambda$CDM&$\Omega_{m0}$ &$0.303^{+0.027}_{-0.036}$&$0.291^{+0.037}_{-0.024}$ &$0.299^{+0.034}_{-0.037}$&$0.303^{+0.027}_{-0.035}$\\
\rule[-2ex]{0pt}{2.5ex}&$\Omega_{r0}$ &$0.000047^{+0.000011}_{-0.000016}$ & $0.000036^{+0.000017}_{-0.000011}$ &$0.000044^{+0.000011}_{-0.000017}$&$0.000041^{+0.000012}_{-0.000017}$\\
\rule[-2ex]{0pt}{2.5ex}& $\Omega_{\Lambda0}$&$0.702^{+0.032}_{-0.033}$& $0.689^{+0.039}_{-0.031}$ &$0.708^{+0.032}_{-0.040}$&$0.694^{+0.042}_{-0.031}$\\
\hline
\end{tabular}}
\label{table4}
\vspace{3pt}
\small{$^*$Anisotropic bumblebee}
\end{table}
\end{center}
%%%%%%%%%%%%%%%%%%%

With these considerations, we have plotted one-dimensional and two-dimensional
marginalized confidence regions ($68\%$ and $95\%$ confidence levels) for
anisotropic bumblebee model parameters $H_0$, $\Omega_{mo}$,
$\Omega_{\Lambda0}$, $\eta$ and $l$ for $H(z)$ (DS-A), $H(z)+ \text{Pantheon
plus}$ (DS-B), $H(z) + \text{Pantheon plus} + \text{BAO}$ (DS-C) and
$H(z) + \text{Pantheon plus} + \text{BAO} + \text{CMB}$ (DS-D) datasets as
shown in Fig.~\ref{fig1}.

Table \ref{table4} shows the constraints ($68\%$ and $95\%$ confidence
 level) on the anisotropic bumblebee model parameters along with the
$\Lambda$CDM model parameters from the different available datasets. From Table
\ref{table4} and Fig.~\ref{fig1}, we found that the tightest constraint can
be obtained from the DS-D dataset, i.e.~joint dataset of $H(z) + 
\text{Pantheon plus} + \text{BAO} + \text{CMB}$ on all the cosmological
parameters for both the anisotropic bumblebee model and the $\Lambda$CDM model.

\begin{figure}[!h]
\centerline{
  \includegraphics[scale = 0.5]{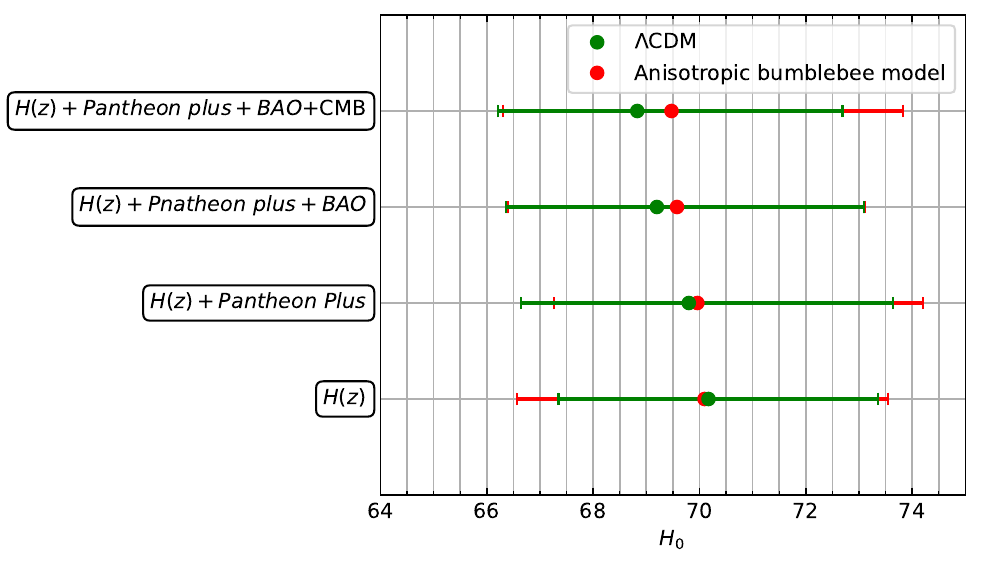}\hspace{0.05cm}
  \includegraphics[scale = 0.5]{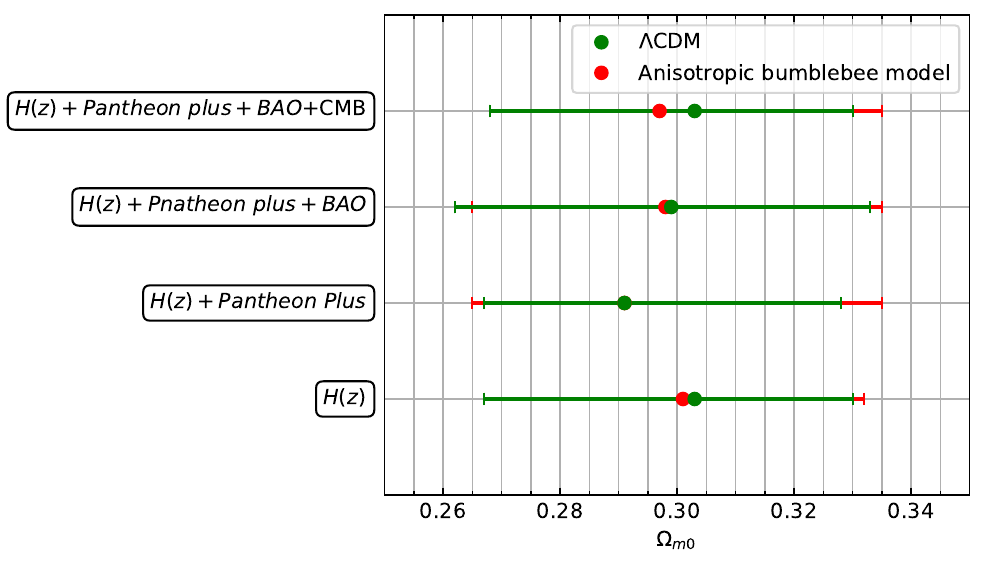}
 }
 \centerline{
  \includegraphics[scale = 0.5]{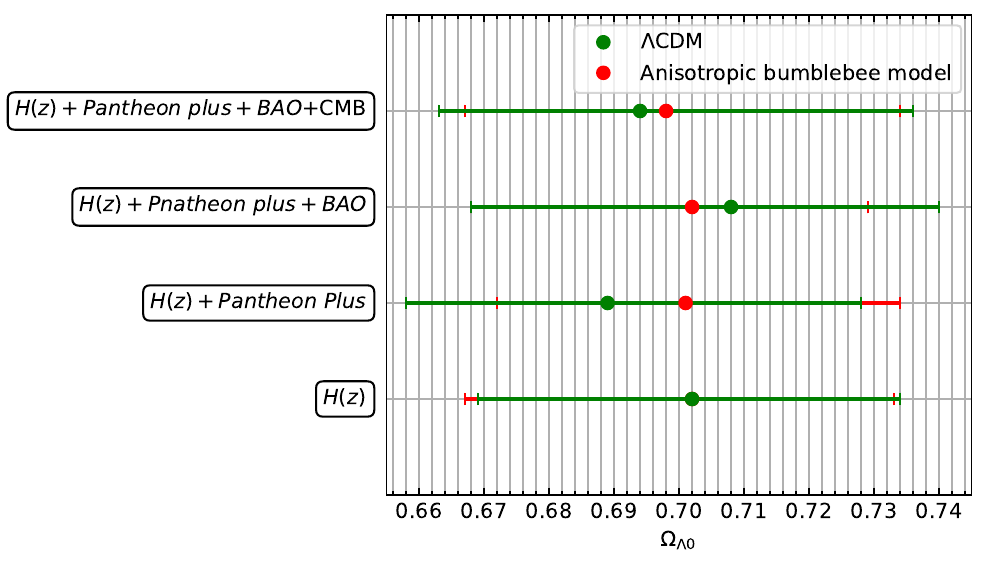}\hspace{0.05cm}
  \includegraphics[scale = 0.5]{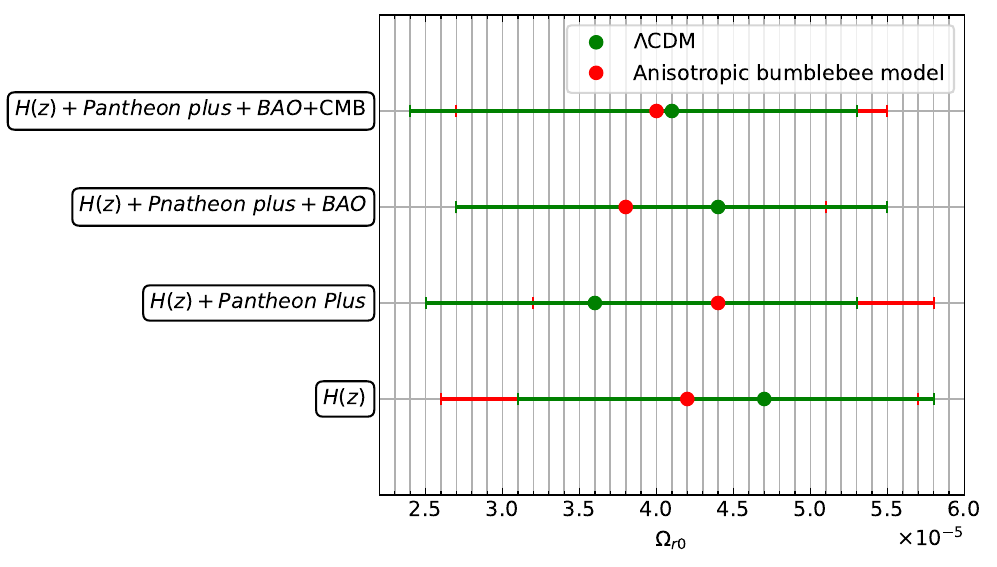}
 }
\vspace{-0.2cm}
\caption{$68\%$ confidence intervals of $H_0,~ \Omega_{m0},~ 
\Omega_{\Lambda 0}$ and $\Omega_{r0}$ for the anisotropic bumblebee model in 
comparison with the $\Lambda$CDM model.}
\label{fig2}
\end{figure}

With the use of Table \ref{table4}, we have tried to compare the $H_0$, 
$\Omega_{m0}$, $\Omega_{\Lambda0}$ and $\Omega_{r0}$ parameters for both the 
models for different dataset combinations within $68\%$ confidence intervals 
as shown in Fig.~\ref{fig2}. The shift of the parameters' values from the 
standard $\Lambda$CDM model due to anisotropic background and bumblebee field 
for different combinations of dataset are clearly observed in the plots of these
figures. The largest deviations of the cosmological parameters from DS-A, DS-B, DS-C and DS-D dataset  are compiled in Table \ref{table5} for both the standard 
$\Lambda$CDM model and the anisotropic bumblebee model. From this Table, we 
can conclude that the deviations are higher in the $\Lambda$CDM model in 
comparison to the anisotropic bumblebee model.
\begin{center}
\begin{table}[!h]
\caption{$68\%$ confidence level maximum deviations of cosmological parameters 
due to different combinations of datasets for the $\Lambda$CDM model and 
anisotropic bumblebee model.}
\vspace{8pt}
\scalebox{1}{
\begin{tabular}{ccccc}
\hline 
\rule[0.1ex]{0pt}{2.5ex}Model \hspace{0.25cm} &\hspace{0.25cm} $\Delta H_0$ \hspace{0.25cm}  & \hspace{0.25cm} $\Delta \Omega_{m0} $ \hspace{0.25cm} & \hspace{0.25cm} $\Delta \Omega_{\Lambda0}$ \hspace{0.25cm} &\hspace{0.25cm} $\Delta \Omega_{r0}$ \hspace{0.25cm}\\ 
\hline
\rule[0.1ex]{0pt}{2.5ex}Anisotropic bumblebee& {0.621} &{0.010} &{0.004} &{0.000006}\\
%\hline
\rule[0.1ex]{0pt}{2.5ex}$\Lambda$CDM& 1.341  &0.012&0.008  &0.000011\\
\hline
\end{tabular}}
\label{table5}
\end{table}
\end{center}
\begin{figure}[!h]
\centerline{
  \includegraphics[scale = 0.65]{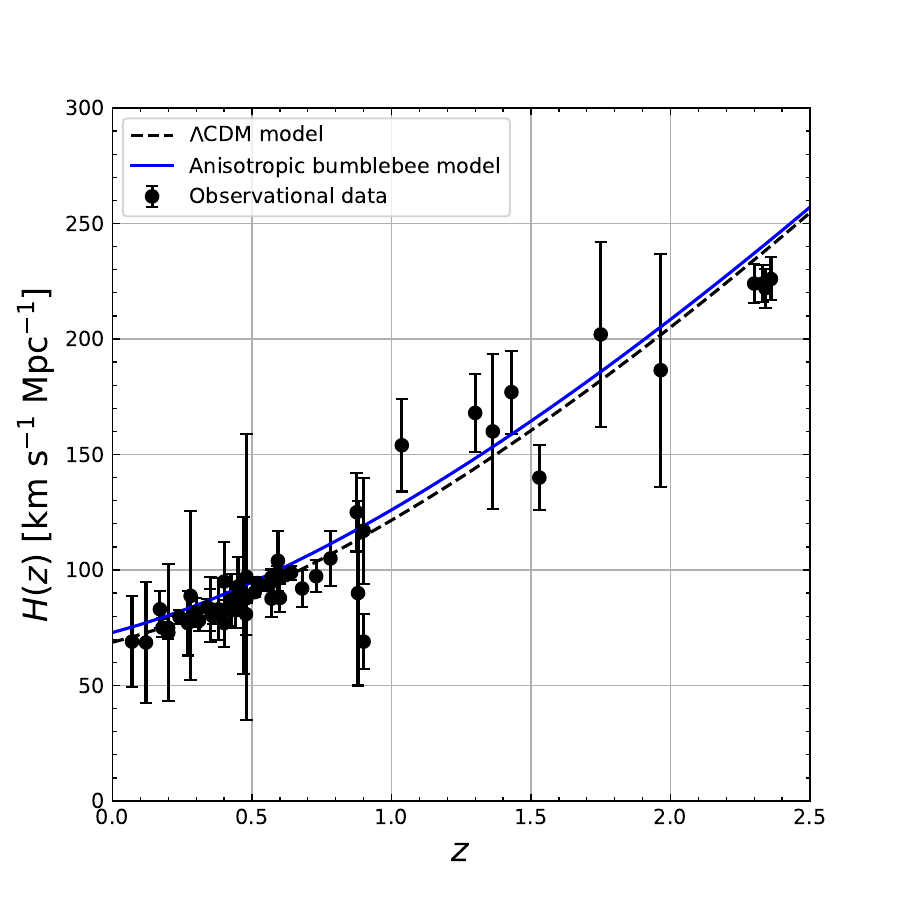}\hspace{0.25cm}
 }
\vspace{-0.3cm}
\caption{Variation of Hubble parameter against cosmological redshift for the 
both $\Lambda$CDM model and anisotropic bumblebee model with the 
constraining set of model parameters along with the observational data.}
\label{fig3}
\end{figure}
\begin{figure}[!h]
\centerline{
  \includegraphics[scale = 0.6]{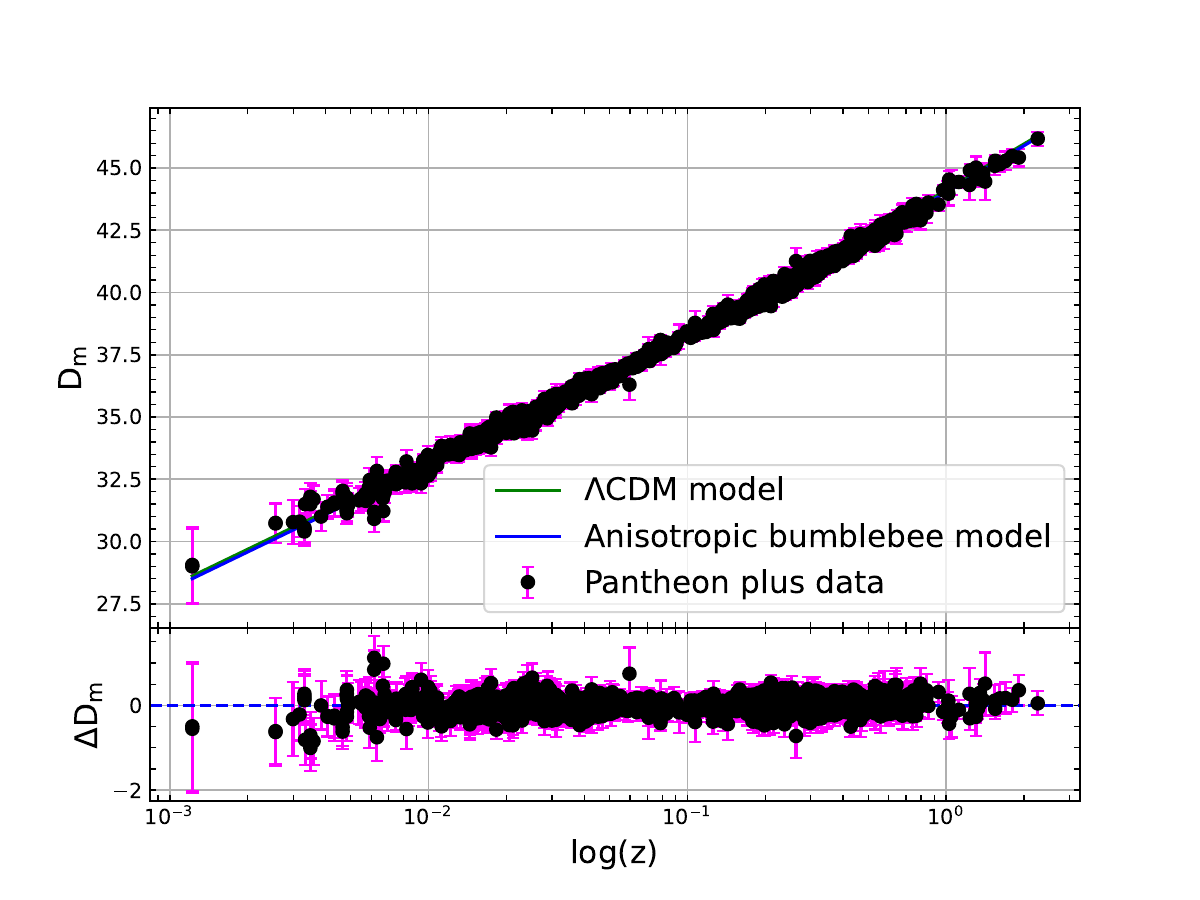}\hspace{0.25cm}
 }
\vspace{-0.2cm}
\caption{Top panel: The Pantheon plus ``Hubble diagram" showing the distance
modulus $D_m$ versus redshift $z$ along with the $\Lambda$CDM and anisotropic
bumblebee results. Bottom panel: Distance modulus residue against cosmological
redshift for the Pantheon plus data relative to the anisotropic bumblebee
model.}
\label{fig4}
\end{figure}

Moreover, we have tried to compare the Hubble parameter versus cosmological 
redshift variations for both models with the parameters constrained for 
the DS-D dataset ($H(z) + \text{Pantheon plus} + \text{BAO} + \text{CMB}$ 
dataset) found from the Table \ref{table4} as shown in Fig.~\ref{fig3}. The 
plot shows that for the estimated values of cosmological parameters, the 
Hubble parameter is consistent with the observational data. However, the 
anisotropic bumblebee model shows deviations from the standard $\Lambda$CDM 
plot with the increase of cosmological redshift $z$. Thus, from this figure, 
we have found that the expansion rate of the early Universe for the 
anisotropic bumblebee model is slower in comparison to that for the standard 
$\Lambda$CDM model. Similarly, we have plotted the distance modulus ($D_m$) 
against cosmological redshift $z$ in Fig.~\ref{fig4} for both $\Lambda$CDM 
and anisotropic bumblebee model along with the distance modulus residue 
relative to anisotropic bumblebee model in the logarithmic $z$ scale for the 
constrained set of model parameters as mentioned in the $H(z)$ vs $z$ plot. 
The plot shows that like $\Lambda$CDM results, the distance modulus for the 
anisotropic bumblebee model is consistent with the observational Pantheon plus 
data obtained from different Type Ia supernovae (SN Ia) for the constrained set 
of model parameters listed in Table \ref{table4}. Furthermore, the plot of 
distance modulus residue also shows that the model is consistent with 
observational data.
\begin{figure}[!h]
\centerline{
  \includegraphics[scale = 0.45]{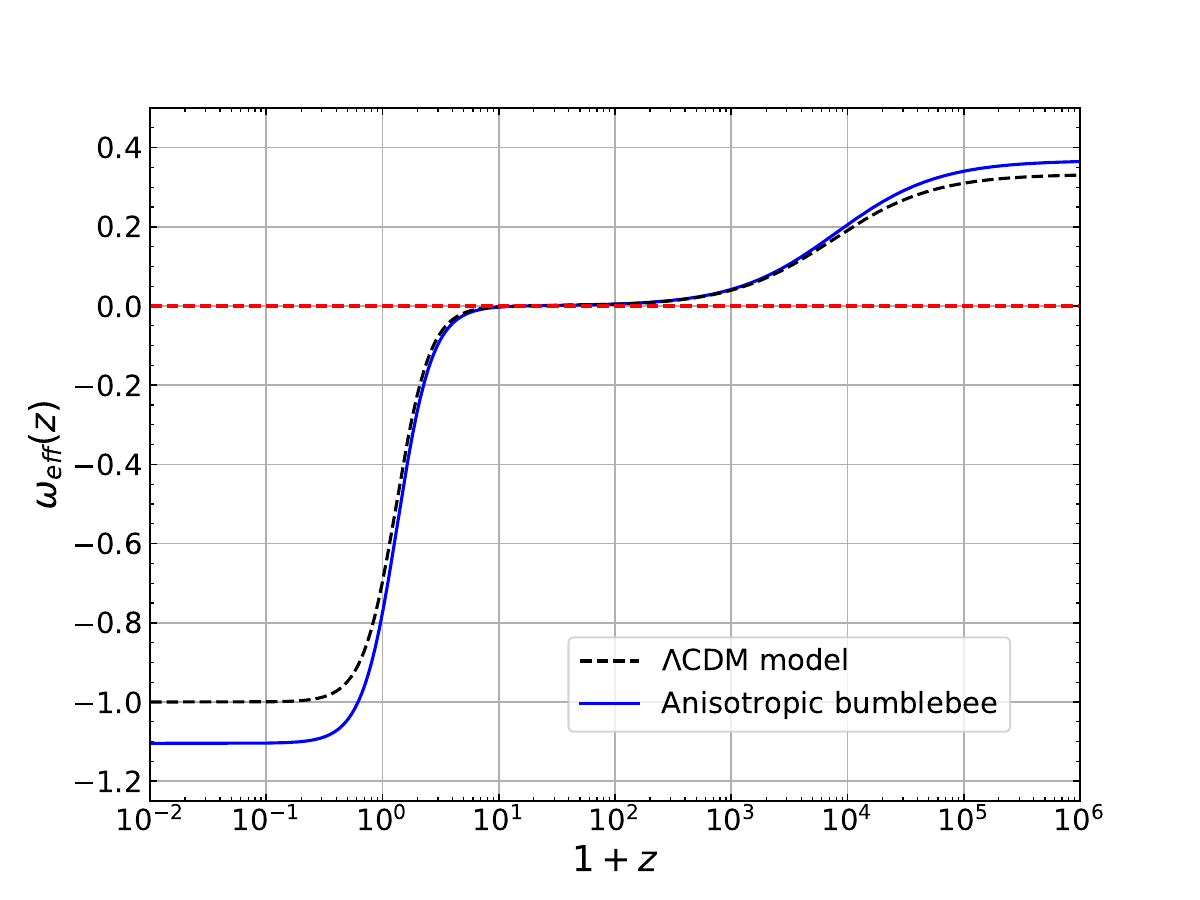}\hspace{0.0cm}
  \includegraphics[scale = 0.45]{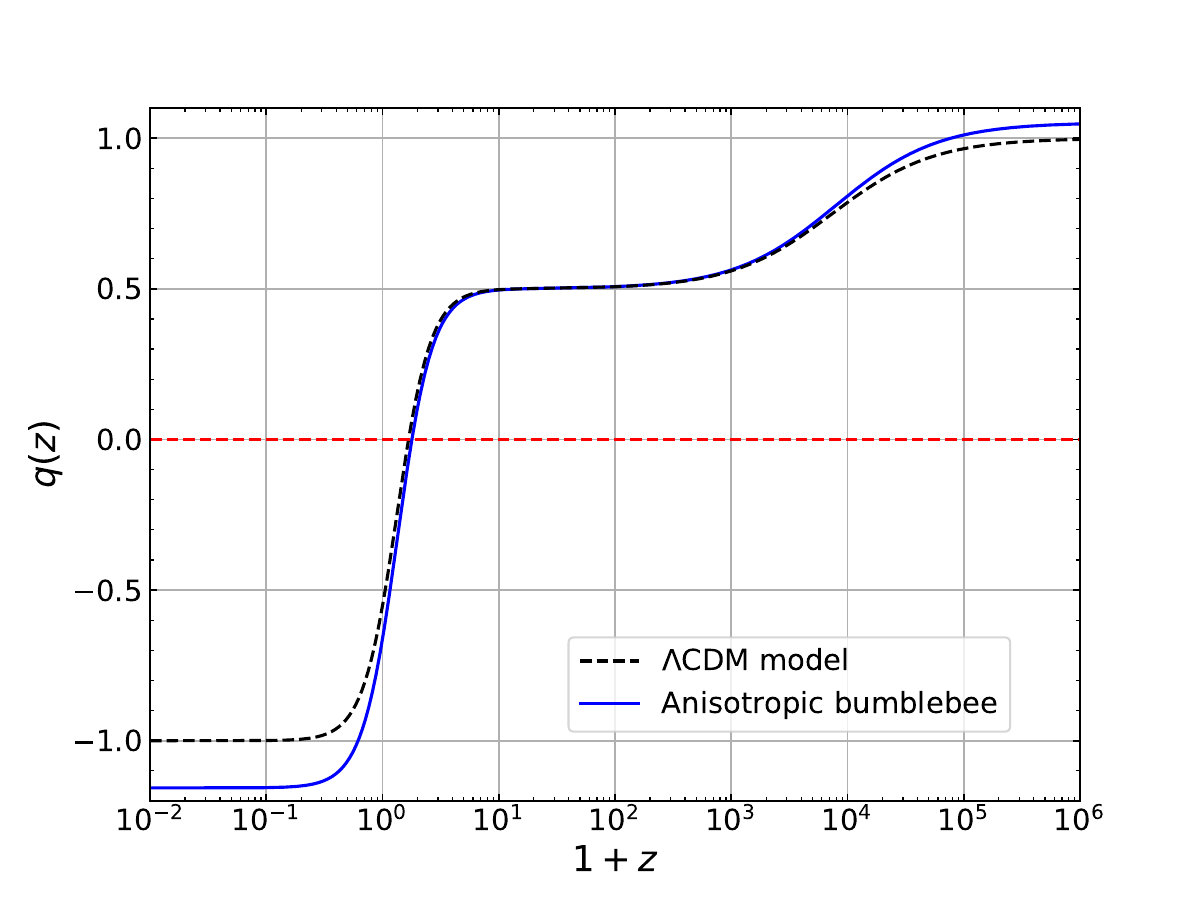}
 }
\vspace{-0.2cm}
\caption{Variation of the effective equation of state $\omega_{eff}$ against 
cosmological redshift (left) and deceleration parameter $q$ against 
cosmological redshift (right) for a constrained set of parameters for both 
$\Lambda$CDM and anisotropic bumblebee models.}
\label{fig4a}
\end{figure}

We have plotted the effective equation of state $\omega_{eff}$ from equation 
\eqref{Om_eff} against cosmological redshift $z$ for the constrained set of 
parameters mentioned in Table \ref{table4} for both $\Lambda$CDM and 
anisotropic bumblebee model in the left panel of Fig.~\ref{fig4a}. The plot 
shows deviations of the anisotropic bumblebee model results from the standard 
$\Lambda$CDM model for {near future and} higher values of $z$, 
indicating the role of 
anisotropy and bumblebee field effect in the early stages of the Universe.
Apart from the $\omega_{eff}$ vs $z$ plot, we have also plotted the 
deceleration parameter $(q)$ using equation \eqref{dec} against the 
cosmological redshift $z$ in the right plot of Fig.~\ref{fig4a} for both 
anisotropic bumblebee and standard $\Lambda$CDM models. In this case also, the 
anisotropic bumblebee model results show deviations from the standard 
$\Lambda$CDM model for {near future and} the higher value of $z$, 
again indicating the role of anisotropy and the effect of bumblebee field in 
the early Universe. However, 
both plots show that anisotropic bumblebee cosmology is consistent with the 
standard cosmology for $z = 0$, i.e. in the present time.

\section{Effect of Anisotropy and bumblebee field on cosmological evolution}\label{6}
\begin{figure}[!h]
\centerline{
  \includegraphics[scale = 0.42]{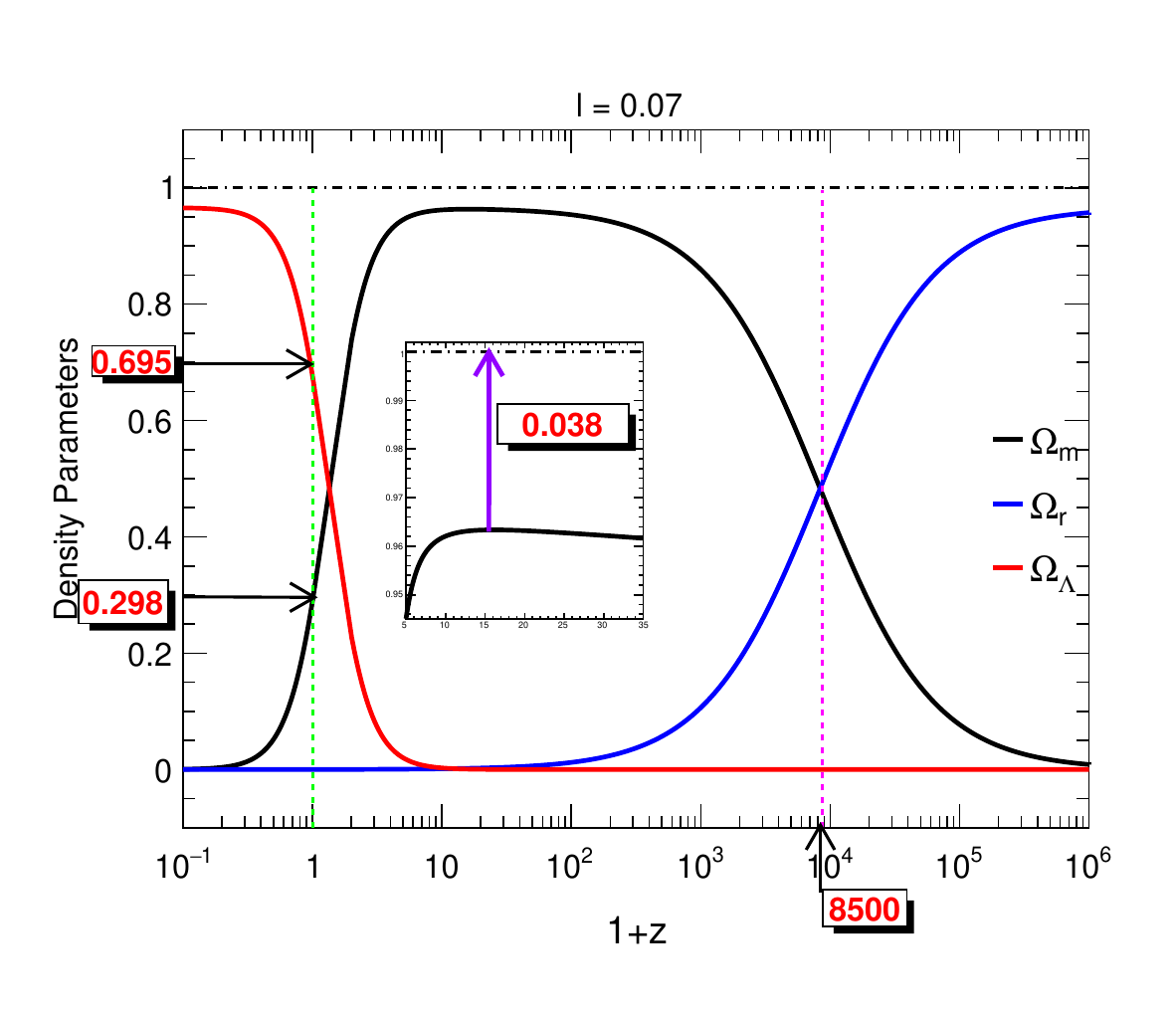}\hspace{0.0cm}
  \includegraphics[scale = 0.42]{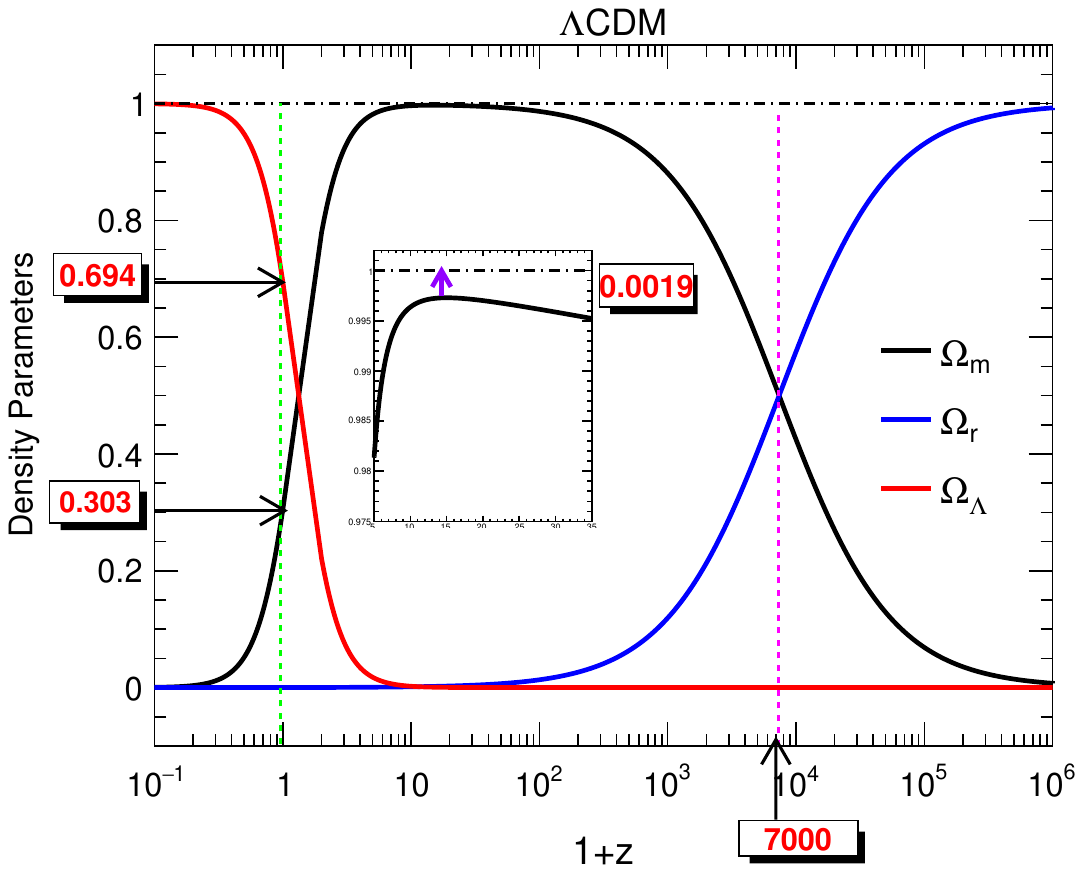}
 }
\vspace{-0.2cm}
\caption{Evolution of density parameters for both anisotropic bumblebee model 
and $\Lambda$CDM model obtained with the constrained set of parameters.}
\label{fig5}
\end{figure}

In this section, we want to investigate how cosmological evolution can be 
affected by a considering bumblebee field in the presence of anisotropy and 
compare it with the standard $\Lambda$CDM results. To this end, we have 
plotted the density parameters of matter ($\Omega_{m}$), radiation 
($\Omega_{r}$) and dark energy ($\Omega_{\Lambda}$) variation with the 
cosmological redshift for both $\Lambda$CDM and anisotropic bumblebee models 
indicating the specific features in the plots as shown in Fig.~\ref{fig5}.

From Fig.~\ref{fig5}, we have found that due to the presence of anisotropy 
and bumblebee field, the value of $(1+z)$ for the transition from 
radiation-dominated to matter-dominated phase in the Universe has been shifted 
from $7000$ of standard $\Lambda$CDM model to $8500$ anisotropic bumblebee 
model. Thus the anisotropic bumblebee model suggests an elongated 
matter-dominated phase. Again, there is a clear sign of an anisotropic effect 
in terms of obtaining the maximum value of density parameters in the 
anisotropic bumblebee model, as its value is much lower than the standard 
$\Lambda$CDM model. In our analysis, we have found that the difference of 
maximum value from unity for $\Omega_{m}$ in $\Lambda$CDM model is
$0.0019$, while in the case of anisotropic bumblebee model, the difference is
$\mathbf{0.038}$. Thus, it indicates that there are some effects of anisotropy and
bumblebee field on the evolution of the matter-dominated as well as other
phases in the Universe. Thus, we can simply say that there are some roles of 
anisotropy and bumblebee fields in cosmic evolution.

\section{Dynamical system analysis of anisotropic bumblebee model}\label{7}
In this section, for a dynamical system analysis of anisotropic bumblebee 
model we have considered the density parameters 
$\Omega_{m} = \frac{\kappa \rho_m}{3H^2}$, 
$\Omega_{r} = \frac{\kappa \rho_r}{3 H^2}$ and 
$\Omega_{\Lambda} = \frac{\Lambda}{3 H^2}$ as dynamical variables and then we 
have rewritten the equations \eqref{FE_v1} and \eqref{FE_v2} as
\begin{eqnarray}
{\Omega_m + \Omega_r + \Omega_{\Lambda} = 1,}\\[5pt]
{\frac{\dot{H}}{H^2} = -\frac{3}{2(1-l)}\left[\frac{1}{3} \Omega_{r} -\Omega_{\Lambda} +\frac{l^2 +(1-l)(2l-1)}{2l-1} \right].}
\end{eqnarray}
Further, we have renamed the parameter $\Omega_m = x$ and $\Omega_r = y$ and 
derived the derivative for each of them with respect to $N = \log{a}$ as
\begin{align}
{\frac{dx}{dN}} & {= \frac{x}{(1-l)}\big[3x  + 4y  - 6+\frac{3l^2 + 3(2l-1)}{(2l-1)}\big],}\\[5pt]
{\frac{dy}{dN} }& {= \frac{y}{(1-l)}\big[3x  + 4y  -7+ \frac{3l^2 + (3+l)(2l-1)}{(2l-1)}\big].}
\end{align}
Moreover, the effective equation of state  and the deceleration parameter can 
further be written as
\begin{align}
{\omega_{eff} }& {= -1 +\frac{1}{(1-l)}\left[\frac{4}{3} y + x -1+\frac{l^2+(1-l)(2l-1)}{2l-1}\right],}\\[5pt]
{q }&{ = -1 +\frac{3}{2(1-l)}\left[\frac{4}{3} y + x -1+\frac{l^2+(1-l)(2l-1)}{2l-1}\right].}
\end{align}
We have compared the critical point analysis for the considered bumblebee model with the standard $\Lambda$CDM results in Table {\ref{table6}}. 
\begin{center}
\begin{table}[h]
\caption{A comparative analysis of the fixed point solutions for anisotropic 
bumblebee model and the $\Lambda$CDM model.}
\vspace{8pt}
\begin{tabular}{|c|c|c|c|c|c|c|c|}
\hline 
\rule[1ex]{0pt}{2.5ex}\hspace{0.3cm} Model \hspace{0.3cm}  & \hspace{0.3cm} $l$ \hspace{0.3cm} & Fixed point & $(x=\Omega_m,y=\Omega_r)$ & Eigenvalues & \hspace{5pt} $\omega_{eff}$\hspace{5pt} & \hspace{0.3cm} $q$ \hspace{0.3cm} & \hspace{0.0cm} Remarks\\[2pt] 
\hline

\rule[1.25ex]{0pt}{2.5ex}Anisotropic &&$P_{a1}$ &${(0,~0.987)}$ & {(1.001,~4.246)} &{0.334}&{1.001}& Unstable point\\ 

\rule[1ex]{0pt}{2.5ex}bumblebee&\textbf{0.07}& $P_{a2}$ &${(0.978,~0)}$  &{(-1.089, 3.065)} &{0.029} &{0.455}& Saddle point\\ 

\rule[1ex]{0pt}{2.5ex}&&  $P_{a3} $ & ${(0,~0)}$  &{(-4.244, -3.244)} & {-1.081}&{-1.220}& Stable point\\

\hline

\rule[1.25ex]{0pt}{2.5ex} &&$P_{b1}$ &$(0,~1)$ & (1,~4) &0.333&1.0& Unstable point\\ 

\rule[1ex]{0pt}{2.5ex}$\Lambda$CDM&0.0& $P_{b2}$ &$(1,~0)$ &(-1,~3) &0.0 &0.5& Saddle point\\ 

\rule[1ex]{0pt}{2.5ex}&&  $P_{b3} $ & $(0,~0)$ & (-4, -3) & -1.0&-1.0& Stable point\\

\hline
\end{tabular}
\label{table6}
\end{table} 
\end{center}
\begin{figure}[!h]
\vspace{-1cm}
\centerline{ 
  \includegraphics[scale = 0.42]{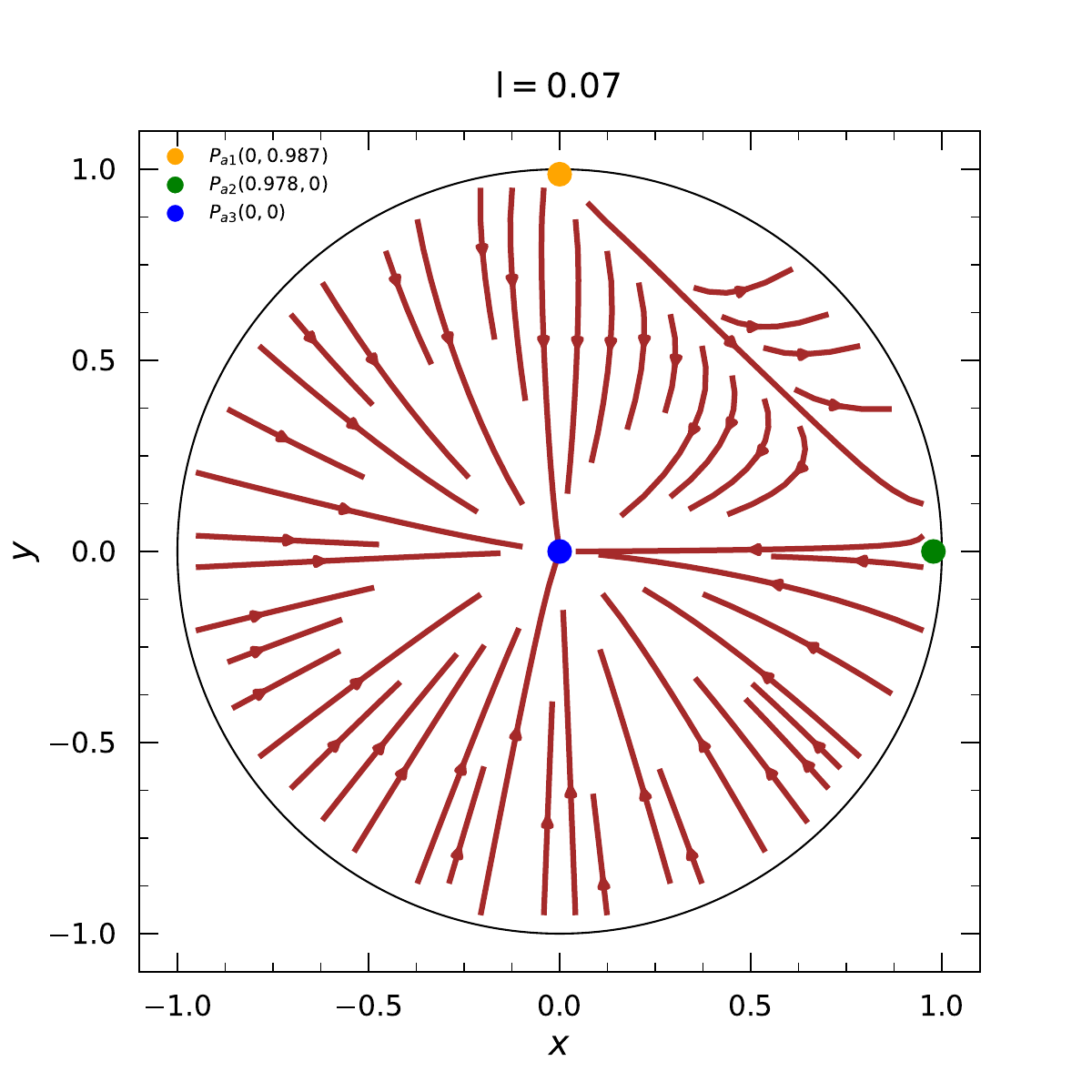}\hspace{0.0cm}
  \includegraphics[scale = 0.42]{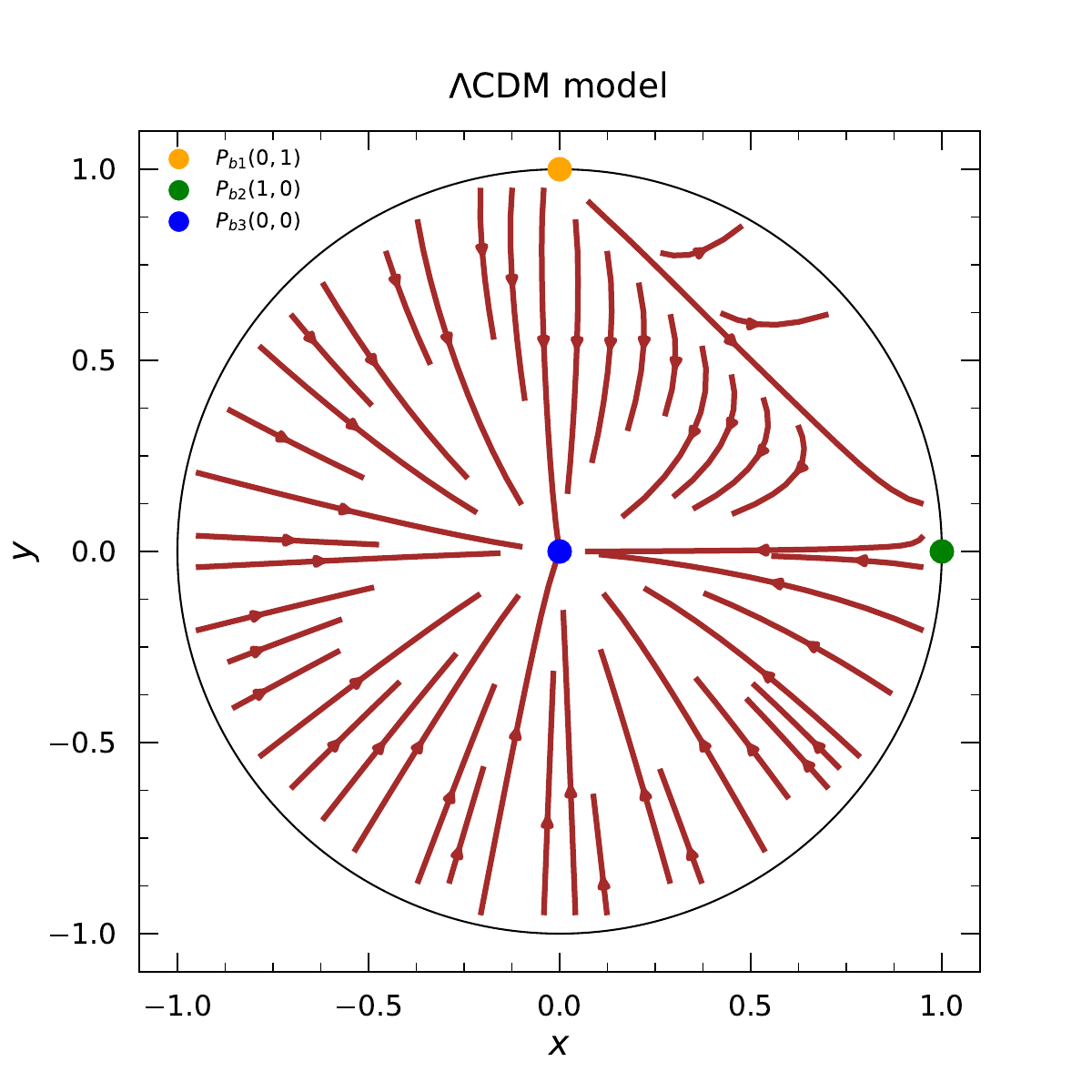}
 }
\vspace{-0.2cm}
\caption{Phase space portraits for anisotropic bumblebee model (left) and 
$\Lambda$CDM model (right).}
\label{fig6}
\end{figure}

The phase space portrait for the anisotropic bumblebee model also shows the 
heteroclinically connected radiation, matter, and dark energy phases as 
predicted by the standard $\Lambda$CDM model (see Fig.~\ref{fig6}). However, 
in contrast to the standard $\Lambda$CDM model, the anisotropic bumblebee 
model shows that the phases have some anisotropic and bumblebee field 
contributions as the critical point solutions $P_{1a}$ and $P_{2a}$ contains 
no exact unity value. Thus, from this analysis, we can say that both 
anisotropy and bumblebee may have some contributions to cosmic evolution.

\section{Conclusions}\label{8}
In this work, we have considered $B_\mu = (B(t),0,0,0)$ bumblebee vector model 
in Bianchi type I Universe and trying to understand its effect on the 
cosmological parameters and cosmological evolutions of the Universe. Further, 
we have considered the Universe as a dynamical system and drawn the phase 
portrait to understand the sequence of different phases of the Universe.

We have started our work by deriving the general form of field 
equations for the considered model and metric as mentioned above and also 
obtained the continuity equations in Section \ref{4}. In the next section, we 
have considered the vacuum expectation value (VEV) condition in the field 
equations of Section {\ref{3}} and also obtained the continuity equations for 
the considered condition. With the help of the field equations, we have further 
obtained the cosmological parameters like the Hubble parameter ($H(z)$), 
luminosity distance ($d_L$), distance modulus ($D_m$), etc., and carry 
forward to study these parameters in our next sections.

In Section \ref{5}, we have constrained our various model parameters 
that have appeared in the cosmological parameters by using the technique of 
Bayesian inference. Here, we have used various data compilations for Hubble 
parameter $H(z)$, Supernovae Type Ia data (Sn Ia), Bao, CMB data, etc.~to 
estimate various model parameters and cosmological parameters which are 
compiled in Table \ref{table4}. This table also includes estimated values of 
cosmological parameters of $\Lambda$CDM model. With these sets of parameters, 
we have compared the values of $H_0$, $\Omega_{m0}$, $\Omega_{r0}$ and 
$\Omega_{\Lambda0}$ for both the models within the $68\%$ confidence interval.
Moreover, we have plotted the Hubble parameter, and distance modulus along 
with distance modulus residue with cosmological redshift for the constrained 
values of model parameters as mentioned above along with the $\Lambda$CDM 
results, and found that the anisotropic bumblebee model shows good agreement 
with standard cosmology and consistent with observational data. However, the 
Hubble parameter shows deviations from the standard $\Lambda$CDM results as 
the $z$ value increases. Furthermore, the effective equation of state and 
deceleration parameters are also plotted against cosmological redshift $z$. 
Both the plots show good agreement with standard cosmology in the current 
scenario while, {for near future and also} for higher $z$, the anisotropic bumblebee model shows 
deviations from the standard results, hence indicating the effect of 
anisotropy and bumblebee field in the early Universe.

In Section \ref{6}, we have studied the effect of anisotropy and 
bumblebee field in various density parameters' evolution. Here, we have 
compared our results with the standard $\Lambda$CDM model and find that there 
is a shift of the transition point from the radiation-dominated to the 
matter-dominated phase of the Universe. In our study, we have found that the 
value of cosmological redshift $(z)$ for the transition from radiation to 
matter phase 
is around $7000$ for $\Lambda$CDM model while for the anisotropic bumblebee 
model, this value shifted to around $8500$. Further, we have noticed that the 
density parameter value never reached unity, hence no pure matter or radiation 
era, but a mixed state of matter, radiation, and dark energy era. Thus, it is
better to say matter-dominated or radiation-dominated state rather than the 
pure state. Also, we have found that the maximum value of these density 
parameters is quite lower than standard $\Lambda$CDM results. Thus there must 
be some role of anisotropy and bumblebee field in the cosmic evolution of the 
Universe.

In Section \ref{7}, we have considered the Universe as a dynamical 
system and hence considered its density parameters as dynamical variables. 
Subsequently, we studied its stability point analysis for both $\Lambda$CDM 
and anisotropic bumblebee models. From this analysis, we have found that the 
anisotropic bumblebee model also shows that there are heteroclinically connected
radiation-dominated, matter-dominated, and dark energy-dominated phases of 
the Universe as suggested by standard cosmology, however, there is a shift of 
critical points from the unity value in anisotropic bumblebee model, which
also confirmed the role of anisotropy and bumblebee field on cosmic evolution 
as mentioned in Section \ref{6}.

{While analyzing our results we have also tried to compare our findings with Ref. \cite{Alam_2017} and \cite{Moresco_2016}. In these works they have explicitly used Baryon Oscillation Spectroscopic Survey (BOSS) data explicitly to constrain the current value of Hubble parameter ($H_0$), angular diameter distance $D_M$ etc. in the isotropic $\Lambda$CDM framework without incorporating any field (like bumblebee or others). Furthermore, these works contain relatively low number of $H(z)$) data ( 30 in Ref. \cite{Alam_2017} ) as compared to our work. Both these works no doubt excellently agree with Planck data release, and gives a very correct knowledge of present scenario of Universe under isotropic framework. Since we are also trying to constrain various cosmological parameters with the help of higher number of $H(z)$ data, Pantheon plus, BAO, CMB data etc. to estimate these parameters, therefore there is a general and fair comparison would distinct our results from the previous works. As mentioned previously, we are considering BI metric along with temporal bumblebee filed , hence the anisotropic characteristics of the spacetime and Lorentz violation parameter also influence our cosmological parameter estimations. For example  the estimated value of $H_0$ in Ref. \cite{Alam_2017} for flat $\Lambda$CDM model is $ 67.6 \pm 0.5 ~km s^{-1} Mpc^{-1}$, while in our case the tightest constraints of $H_0$ for BI metric in presence of temporal bumblebee field is ${69.474^{+4.358}_{-3.167} ~km s^{-1} Mpc^{-1}}$ . Similarly, the estimated values of other cosmological parameters are different from the previous works. Apart from that, with the introduction of bumblebee field and for an anisotropic background provides some additional information like Lorentz violation length $l$, possible anisotropic characteristics etc. of the Universe.}

Finally, in conclusion, we have observed that with the consideration of 
anisotropy and the bumblebee field, the evolution of the Universe i.e. the 
evolution of the various phases of the Universe is somehow affected and hence 
may provide an interesting scenario of cosmic expansion. {However, the deviations of cosmological parameters from standard FLRW cosmology  for considering bumblebee field as a temporal field in our work is not very much at $z=0$ or near past. Similar less deviations from FLRW results has also been stated in Ref.} \cite{Maluf_2021} { where bumblebee field has been considered in spatial direction. Thus we can conclude that irrespective of considering bumblebee either in temporal or spatial direction results anisotropy but the deviations from standard cosmology.} However for studying early stage of the Universe more observational data on the early Universe need to be require which may help us to understand this scenario more clearly in the near future.

\section*{ACKNOWLEDGEMENT}
UDG is thankful to the Inter-University Centre for Astronomy and Astrophysics 
(IUCAA), Pune, India for the Visiting Associateship of the institute.

\bibliographystyle{apsrev}

\end{document}